\documentclass[aps,pre,twocolumn,superscriptaddress,superscriptreference]{revtex4-2}

\usepackage{amsmath,bbold,bm,amssymb,scalerel,mathtools}
\usepackage{graphicx}
\usepackage{color}
\usepackage{enumitem}
\usepackage{algorithm,algpseudocode}
\usepackage{multirow}
\usepackage{colortbl,booktabs}
\usepackage{placeins}
\usepackage[usenames,dvipsnames]{xcolor}
\usepackage{lipsum}

\usepackage[colorlinks, linkcolor=BrickRed, urlcolor=blue!50!black, citecolor=blue!50!black]{hyperref}

\newcommand\equalhat{%
	\let\savearraystretch\arraystretch
	\renewcommand\arraystretch{0.3}
	\begin{array}{c}
		\stretchto{
			\scalerel*[\widthof{=}]{\wedge}
			{\rule{1ex}{3ex}}%
		}{0.5ex}\\ 
		=%
	\end{array}
	\let\arraystretch\savearraystretch
}
\newcommand{\myrowcolour}{\rowcolor[gray]{0.925}}

\newcommand{\diff}{\text{d}}
\newcommand\difft{\text{d} t}
\newcommand\kB{k_{\mathrm{B}}}

\newcommand{\gj}[1]{\textcolor{black}{#1}}

\usepackage[normalem]{ulem}

\begin{document}

\title{Dynamic Coarse-Graining of Linear and Non-Linear Systems: Mori-Zwanzig Formalism and Beyond}

\author{Bernd Jung}
\email{bernd@ralfj.de}
\affiliation{Independent Scholar, Eltville, Germany}

\author{Gerhard Jung}
\email{gerhard.jung.physics@gmail.com}
\affiliation{Laboratoire Charles Coulomb (L2C), Université de Montpellier, CNRS, 34095 Montpellier, France}

\begin{abstract}
	
	To investigate the impact of non-linear interactions on dynamic coarse graining, we study a simplified model system, featuring a tracer particle in a complex environment. Using a projection operator formalism and computer simulations, we systematically derive generalized Langevin equations (GLEs) describing the dynamics of this particle. We compare different kinds of linear and non-linear coarse-graining procedures to understand how non-linearities enter reconstructed generalized Langevin equations and how they influence the coarse-grained dynamics. For non-linear external potentials, we show analytically and numerically that the non-Gaussian parameter and the incoherent intermediate scattering function will not be correctly reproduced by the GLE if a linear projection is applied. This, however, can be overcome by using non-linear projection operators. We also study anharmonic coupling between the tracer and the environment and demonstrate that the reconstructed memory kernel develops an additional trap-dependent contribution. Our study highlights some open challenges and possible solutions in dynamic coarse graining.

\end{abstract}

\maketitle

\section{Introduction}

Chemical and biological soft matter systems are governed by processes on very different length and time scales. Computational studies therefore often rely on coarse graining. This term describes the process of removing microscopic degrees of freedom to find models which are computationally more efficient but still allow for realistic description of static and dynamic properties \cite{mueller2002,noid2013_cg,brini2019_cg,rudzinski2019recent}. While structural properties in coarse-grained (CG) models can be described and reconstructed by effective (pair) potentials \cite{izvekov2005multiscale,reith2003ibi}, the systematic description of dynamic quantities is even more complex and often involves parameterization of a generalized Langevin equation (GLE) \cite{Mori1965,Zwanzig2001}. This introduces additional dissipative and thermal forces to the model, emerging due to the systematic removal of degrees of freedom.

The GLE has been originally derived using the Mori-Zwanzig formalism \cite{Zwanzig1961,Mori1965,Zwanzig2001} and has recently seen a surge in popularity for CG modeling of chemical and biological systems with consistent dynamics \cite{McKinley2009_diffusion,Shin2010,cordoba2012elimination,gottwald2015_gle,wu2018_adhesion,meyer2021_gle,klippenstein2022_asakura,Ayaz2022_GLENonLinear,Vroylandt2022_nonlinear,widder2022noise}, including polymer melts \cite{karniadakis2017_melt}, colloidal suspensions \cite{jung2018_GLD} and water \cite{post2022molecular,klippenstein2023_water} as reviewed in Ref.~\cite{klippenstein2021introducing}. While these works highlight the general applicability and generalizability of the GLE, several fundamental properties and consequences of the GLE are still debated. This includes particularly the inclusion of non-linearities. While the original projection formalism using the Zwanzig projector is a non-linear theory \cite{Zwanzig1961}, the derivation of the GLE requires linear projections introduced by Mori \cite{Mori1965}, leading to a purely linear equation. For many applications in physical systems, however, non-linearities, in particular non-linear potentials of mean force, have been reintroduced \cite{karniadakis2017_melt,jung2018_GLD,Ayaz2022_GLENonLinear,post2022molecular,klippenstein2023_water}. Consequently, over the years much work has been invested into reintroducing non-linearities to the projection operator formalism and the GLE \cite{zwanzig1973nonlinear,Kinjo2007,Xing2009_nonlinear,Cairano2021_nonlinear}, which has been intensified in the last year \cite{Glatzel_2021,Ayaz2022_GLENonLinear,Vroylandt2022_nonlinear,Vroylandt_2022,SciPostPhys.13.4.078}. Due to these great efforts we now have a good theoretical understanding of how non-linearities can be included into the GLE. The practical consequences for dynamic coarse graining are, however, not yet widely explored. This concerns, in particular, the question how non-linearities in the underlying system influence the form and dynamics of the CG model.

In this work we will bridge this research gap and investigate a simplified system \gj{in thermal equilibrium} using analytical theory, linear and non-linear projection operator formalisms and computer simulations for a detailed investigation into the impact of non-linearities on dynamic CG models. The system consists of a tagged particle, trapped in an external potential and coupled to damped stochastic oscillators, similar, but not identical, to previously suggested systems \cite{CLModel,hanggi2007generalized,Xing2009_nonlinear,Ayaz2022_GLENonLinear}. The aim of our study is to describe the dynamics of the tagged particle by a GLE model. We investigate two kinds of non-linearities: one connected to the external potential and one connected to the particle-oscillator coupling. Importantly, we study the dynamics not only using two-point autocorrelation functions, as is commonly done, but we investigate also higher-order correlation functions, including the non-Gaussian parameter and the incoherent inherent scattering function. We find that these more complex descriptors are oftentimes not properly reproduced by CG models albeit the dynamics seems correct on the level of two-point functions.

Our manuscript is organized as follows. In Sec.~\ref{sec:SNCL} we will introduce the model which we analyze throughout this work. Furthermore, we present several specific analytical results and will apply a linear and a non-linear projection operator formalism to the system to derive CG equations of motion. Sec.~\ref{sec:simulations} describes the numerical procedures and computer simulations performed to accompany the analytical results derived in the previous section. We present our findings for three variations of the model in Sec.~\ref{sec:linear} (purely linear), Sec.~\ref{sec:anharmonic_trap} (anharmonic trap) and Sec.~\ref{sec:anharmonic_coupling} (anharmonic coupling). We discuss the implications for dynamic coarse graining of general soft matter systems in Sec.~\ref{sec:discussion}. We summarize and conclude in Sec.~\ref{sec:conclusion}.

\section{Stochastic Non-Linear Caldeira-Leggett (SNCL) Model }
\label{sec:SNCL}

In this paper we analyze a simplified, one-dimensional model for the interaction of a tagged particle with its environment. The model is described by the position $x_0$ and velocity $v_0$ of the particle, as well as the positions $x_i$ and velocities $v_i$, $i=1,\ldots, N$ of the $N$ oscillators describing the environment. The equations of motion are given by
 	\begin{align}\label{eq:eom}
\frac{\diff}{\difft} x_0 &= v_0\\
m_0\frac{\diff}{\difft} v_0 &= F_0(t) = - V_{\text{e}}^\prime(x_0(t)) +  \sum_{i=1}^N V_i^\prime (\Delta x_i(t)) \nonumber\\
\frac{\diff}{\difft} x_i &= v_i \nonumber\\
m_i\frac{\diff}{\difft} v_i &= - V_i^\prime (\Delta x_i(t)) - \gamma_i v_i(t) + W_i(t) \nonumber
\end{align}
with external potential $V_{\text{e}}(x_0(t))$, $\Delta x_i(t) = x_0(t) - x_i(t)$, coupling potential between particle and oscillators $V_i(\Delta x_i(t))$, friction coefficient $\gamma_i$ and white noise $\langle W_i(t) W_i(t^\prime) \rangle = 2 \gamma_i \kB T \delta(t-t^\prime)$. \gj{The steady-state dynamics of the model described by Eq.~\ref{eq:eom} is in thermal equillibrium and samples the canonical ensemble. } The model is strongly related to the classical version of the Caldeira-Leggett model \cite{CLModel,hanggi2007generalized} with the addition of stochastic interactions and generalized interaction potentials. In the remainder of this manuscript we will therefore refer to it as the stochastic non-linear Caldeira-Leggett (SNCL) model. For linear potentials, the model is equivalent to the analytically solvable model recently studied in Ref.~\cite{Ayaz2022_GLENonLinear} and it also shares some similarities with the dissipative, analytically solvable model studied in Refs.~\cite{doerries2021correlation,Jung_2022}. \gj{As a physical interpretation of the model one could see the ``oscillators'' as the first solvation shell of a molecule or colloid  with which it interacts directly. The remaining fluid is represented as a bath in thermodynamic limit via the dissipative and stochastic interactions acting on the oscillators.}

If not otherwise stated, we use $\kB T = 1 \epsilon$, $ m_0 = 1 m$, $\gamma_i = 1.5 m \tau^{-1}$, and $m_i = 0.2 m$. This also defines the units of energy $\epsilon$, mass $m$, time $\tau$, and length $\sigma = \tau \sqrt{\epsilon / m}$.

In the following, we will investigate three different variations of the SNCL model:
\begin{itemize}
	\item \textbf{Linear}: The linear version is defined by the harmonic potentials $V_{\text{e}}(x_0) = \frac{1}{2} a_{\text{e}} x_0^2$ and $V_i(\Delta x_i) = \frac{1}{2} a_i (\Delta x_i)^2$ and is fully analytically solvable \cite{Ayaz2022_GLENonLinear}.
	\item \textbf{Anharmonic trap}: In this variation we define a non-linear trap $V_{\text{e}}(x_0) = \frac{1}{2} a_{\text{e}} x_0^2 + \frac{1}{4} b_{\text{e}} x_0^4$ but keep the harmonic coupling $V_i(\Delta x_i) = \frac{1}{2} a_i (\Delta x_i)^2$. This model is analytically solvable for very small $b_{\text{e}}$ and for limiting values of the dynamics (see Sec.~\ref{sec:limit}). 
	\item \textbf{Anharmonic coupling}: Finally, we also study the effect of non-linear coupling between particle and oscillators $V_i(\Delta x_i) = \frac{1}{2} a_i (\Delta x_i)^2+ \frac{1}{4} b_i (\Delta x_i)^4$, where $b_i$ defines the strength of the non-linearities. Analytical results are only available in specific limits $t=0$ and $t \rightarrow \infty$ (see Sec.~\ref{sec:limit}).  \gj{We will use a linear trap $V_{\text{e}}(x_0) = \frac{1}{2} a_{\text{e}} x_0^2$.}
\end{itemize}

\subsection{The generalized Langevin equation (GLE)}\label{sec:GLES}

In this work, we will derive CG equations of motion for the tagged particle by integrating out, or projection out, the coupled oscillators.  We will focus on two different GLEs.

The first is the Mori-Zwanzig (MZ) GLE:
\begin{align}\label{eq:mz_gle}
m\frac{\diff}{\difft} v_0 = - \tilde{a}_{\text{e}} x_0 - \int_{0}^{t} \text{\diff} s K_\text{MZ}(t-s) v_0(s) + \eta_\text{MZ}(t)
\end{align}
with effective harmonic strength $\tilde{a}_{\text{e}} = \frac{\kB  T}{ \langle x_0 ^2 \rangle}$, MZ memory kernel $K_\text{MZ}(t)$ and thermal fluctuations $\eta_\text{MZ}(t)$. The MZ GLE is the outcome of the Mori-Zwanzig projection operator formalism \cite{Zwanzig1961,Mori1965,Zwanzig2001}, which is purely based on linear projections and is thus inherently linear. This process of projecting non-linear interactions onto linear fluid transport coefficients is sometimes called ``fluctuation renormalization'' \cite{Zwanzig2001}. We interpret the $\eta_\text{MZ}(t)$ as a stochastic process fulfilling the second fluctuation-dissipation theorem (2FDT) $\langle \eta_\text{MZ}(t)\eta_\text{MZ}(0)\rangle = \kB  T K_\text{MZ}(t)$, which is very natural for our model since we start the coarse graining already from a stochastic differential equation \cite{zhu2021_stochasticgle}  and not from Hamiltonian dynamics as assumed for the original MZ formalism.

We will also investigate the non-linear (NL) GLE:
\begin{align}\label{eq:nl_gle}
m\frac{\diff}{\difft} v_0 = - V_{\text{e}}^\prime(x_0) - \int_{0}^{t} \text{\diff} s K_\text{NL}(t-s) v_0(s) + \eta_\text{NL}(t)
\end{align}
which differs from the above MZ GLE only by featuring the full non-linear external potential $V_{\text{e}}(x_0).$ The NL GLE can be derived using a non-linear projection operator formalism \cite{Vroylandt2022_nonlinear} (see Sec.~\ref{sec:MZ} for details) or by analytically integrating out the oscillator, as will be shown in the following.

If the particle only couples harmonically to the oscillators, it is possible to integrate out the oscillator equations of motion and thus derive closed forms for the dynamics of the tagged particle (see Appendix \ref{app:analytic_bo0} for details). The resulting equations of motion are identical to the above NL GLE with memory kernel
 \begin{equation}\label{eq:MemoryAnalytic}
K_\text{NL}(t) = \sum_{i=1}^{N} a_i e^{-\delta_i t} \left[  \cos(\tilde{\omega}_i t) + \frac{\delta_i}{\tilde{\omega}_i}  \sin(\tilde{\omega}_i t)  \right]
\end{equation}
where $ \delta_i = \frac{\gamma_i}{2 m_i} $, $\omega_i = \sqrt{a_i/m_i}$, and $ \tilde{\omega}_i = \sqrt{\omega_i^2 - \delta_i^2 } $. Here, we have assumed that the oscillators are underdamped (i.e., $ \omega_i > \delta_i $). See Appendix \ref{app:analytic_bo0} for critical damping and overdamping. Integration yields $\int_{0}^{\infty} \difft K_\text{NL}(t) = \sum_{i=1}^{N} \gamma_i$ in all cases.

For the autocorrelation of thermal fluctuations we find
\begin{equation}\label{eq:FDTAnalytic}
C_\eta(t)=\langle \eta_\text{NL}(t) \eta_\text{NL}(0) \rangle = \kB  T K_\text{NL}(t).
\end{equation}

In the following subsections \ref{sec:limit} and \ref{sec:MZ} we will present specific analytical results for these two GLE models and several properties characterizing the dynamics of the systems. 

\subsection{Analytical results for limiting cases}
\label{sec:limit}

While analytical calculations for general anharmonic systems are very difficult, we will derive in the following exact results for specific quantities and limits, such as instantaneous fluctuations ($t=0$) and long-time behavior ($t\rightarrow \infty$).

\subsubsection{Effective harmonic strength $\tilde{a}_{\text{e}}$}

The effective harmonic strength $\tilde{a}_{\text{e}}= \frac{\kB  T}{ \langle x_0^2 \rangle }$ was introduced for the MZ GLE and requires the calculation of the phase space integral
\begin{equation}
\langle x_0^2 \rangle = \frac{1}{Z} \int_{- \infty}^{\infty} \diff x_0 \, x_0^2 e^{-\beta (\frac{1}{2} a_{\text{e}} x_0^2 + \frac{1}{4} b_{\text{e}} x_0^4)}
\end{equation}
with $Z = \int_{- \infty}^{\infty} \diff x_0 e^{-\beta (\frac{1}{2} a_{\text{e}} x_0^2 + \frac{1}{4} b_{\text{e}} x_0^4)}$ and inverse temperature $\beta^{-1} = \kB  T$.
While in the general case we rely on numerical integration of this one-dimensional integral, we can analytically solve some special cases (cf. Appendix \ref{app:atilde}):
\begin{itemize}
	\item For $b_{\text{e}} = 0$ we find, of course, $\tilde{a}_{\text{e}} = a_{\text{e}}$.
	\item For $a_{\text{e}}=0$ it is $\tilde{a}_{\text{e}}= \frac{ \Gamma(0.25)^2}{\sqrt{8} \pi} \sqrt{b_{\text{e}} \kB  T} $, with the Gamma function $\Gamma(x)$.
	\item To linear order in $b_{\text{e}}$ we find for small $b_{\text{e}}$, $\tilde{a}_{\text{e}}= a_{\text{e}} + \frac{3 b_{\text{e}} \kB  T}{a_{\text{e}}} + \mathcal{O}(b_{\text{e}}^2)$.
	\item Generally, $\tilde{a}_{\text{e}} = \frac{a_{\text{e}}}{2} + \sqrt{\frac{a_{\text{e}}^2}{4} + \theta b_{\text{e}} \kB  T}$, with $\theta = \frac{\langle x_0^4\rangle}{\langle x_0^2\rangle ^2}$. $\theta$ is bounded from below by  $\theta \geq \frac{ \Gamma(0.25)^4}{{8} \pi^2} \approx 2.18844$, which emerges for $b_{\text{e}}/a_{\text{e}}^2 \rightarrow \infty$, and from above by $ \theta \leq 3 $ for $b_{\text{e}}/a_{\text{e}}^2 \rightarrow 0$.   
\end{itemize}

\subsubsection{Mean squared displacement and non-Gaussian parameter}
\label{sec:observables}

Two of the dynamical properties that will be studied within this work are the mean squared displacement $\text{MSD}(t) = \langle (x_0(t) - x_0(0))^2 \rangle$ and the non-Gaussian parameter $\alpha(t) = \frac{1}{3} \frac{\langle (x_0(t) - x_0(0))^4 \rangle}{\text{MSD}(t)^2}  -1$. The latter describes, as the name suggests, deviations from pure Gaussian displacements on the time scale $t$ and is intensively used in glass physics \cite{Flenner2005_nonGaussian} and biophysics \cite{wang2012_nonGaussian,cherstvy2019_nonGaussian} to characterize dynamic heterogeneities.

Both quantities are zero at $t=0$ per definition, and in the limit $t\rightarrow \infty$ they approach to
\begin{align}\label{eq:msd}
\lim\limits_{t\rightarrow \infty }\text{MSD}(t) &= 2 \langle x_0^2 \rangle = 2 \frac{\kB  T}{\tilde{a}_{\text{e}}}
\end{align}
\begin{align}\label{eq:alpha}
\lim\limits_{t\rightarrow \infty }\alpha(t) &= \frac{\theta}{6} - \frac{1}{2}
\end{align}
which is completely independent of the particle-oscillator coupling and only depends on the trapping itself.

\subsubsection{Incoherent intermediate scattering function (ISF)}

We finally also calculate the ISF, defined as $F_{\text{s}}(q,t) = \langle \cos (q [x_0(t) - x_0(0)]) \rangle$, which is an important experimental quantity measurable via scattering experiments \cite{vanMegen1998_isf}. The wave number $q$ in this case represents the inverse length scale of the displacements probed in the scattering experiment.

For small $q$ we can expand the ISF to derive the well known relation $F_{\text{s}}(q,t) = {1 -} \frac{q^2}{2} \text{MSD}(t) + \frac{q^4}{8}\text{MSD}(t)^2 (\alpha(t) + 1) + \mathcal{O}(q^6) $, which connects the ISF to the previously introduced dynamic quantities MSD and non-Gaussian parameter. In the limit $t\rightarrow \infty$ we can assume that $x_0(t)$ and $x_0(0)$ are independent and thus calculate the ISF via the integral
\begin{equation}\label{eq:ISF_limit}
F_{\text{s}}(q,t\rightarrow \infty) = \frac{1}{Z^2} \int_{- \infty}^{\infty} \diff x \int_{- \infty}^{\infty} \diff y \, \cos (q (x-y)) e^{-\beta V_{\text{e}}(x)} e^{-\beta V_{\text{e}}(y)}
\end{equation}
which we evaluate using numerical integration.

\subsection{Non-linear projection operator formalism}
\label{sec:MZ}

Non-linear GLEs have been used since several years to describe complex dynamics of soft matter systems \cite{PhysRevE.75.051109,karniadakis2017_melt,jung2018_GLD}. Particularly in the past two years it has, however, been controversially discussed whether and how such NL GLEs can be derived systematically using projection operators \cite{Glatzel_2021,Ayaz2022_GLENonLinear,Vroylandt2022_nonlinear}. Here, we will apply the methods presented in Refs.~\cite{Vroylandt2022_nonlinear,zhu2021_stochasticgle} to derive the NL GLE using non-linear projection operators. Readers who are not interested in these more technical and analytical details, which are not necessary for the understanding of the main results, can immediately jump to Sec.~\ref{sec:simulations} for details on the simulation techniques.

\subsubsection{Time-evolution operator}

Different from Ref.~\cite{Vroylandt2022_nonlinear} the microscopic dynamics of our system is stochastic. We will therefore follow the lines of Refs.~\cite{zhu2021_stochasticgle,Jung_2022} and apply the projection operator formalism on noise-averaged observables $ \varphi(\bm{X}(t)) = \mathbb{E}_{\bm{W}(t)}\left[ \phi(\bm{X}(t)) | \bm{X}(0) \right] $, where $\phi(\bm{X}(t))$ describes an arbitrary observable defined on the phase space variables $\mathbf{X}(t) = \{x_0(t),v_0(t), \ldots, x_N(t), v_N(t) \}$. For these noise-average observables it is possible to write the time-evolution of the system in the form
\begin{equation}
 \varphi(\bm{X}(t)) = e^{t \mathcal{K}} \varphi(\bm{X}(0))
\end{equation} 
with the Fokker-Planck operator \cite{zhu2021_stochasticgle}
 	\begin{align}\label{key}
\mathcal{K} &= v_0 \frac{\partial }{\partial x_0}  - V_{\text{e}}^\prime(x_0) \frac{\partial }{\partial v_0} + \sum_{i>0} V_i^\prime (\Delta x_i) \frac{\partial }{\partial v_0}\nonumber\\ &+ \sum_{i>0} v_i \frac{\partial }{\partial x_i} - \sum_{i>0} V_i^\prime (\Delta x_i) \frac{\partial }{\partial v_i} - \frac{\gamma_i}{m_i} v_i \frac{\partial }{\partial v_i} \nonumber\\&+ T \sum_{i>0} \frac{\gamma_i}{m_i^2} \frac{\partial^2 }{\partial v_i^2}.
\end{align}
Having introduced the time-evolution as described above, we can adapt the notation and the full formalism of Ref.~\cite{Vroylandt2022_nonlinear} with the only change of replacing the Liouville operator $\mathcal{L}$ by $\mathcal{K}$. \gj{In Appendix \ref{app:MZ} we give a condensed recapitulation of the formalism, but refer to Ref.~\cite{Vroylandt2022_nonlinear} for the complex details.}

\subsubsection{Non-linear GLE}

\gj{It has been shown in Ref.~\cite{Vroylandt2022_nonlinear} that the non-linear projection scheme can be used to derive the differential equation,}
\begin{align}
\dot{x}_0 &= v_0\\
\dot{v}_0& = f_b(x_0)  - \int_0^t \text{\diff}s K(s,x_0(t-s)) \gj{{v}_0}(t-s) + \eta(t)
\end{align}
including the mean force $f_b(x_0)$, the position-dependent memory kernel $K(s,x_0(t-s))$ and the noise $\eta(t)$. Applying the above projection operator formalism, we can derive explicit expressions for these terms under the assumption that $  V_{\text{e}}(x_0) = \frac{1}{2} a_{\text{e}} x_0^2 + \frac{1}{4} b_{\text{e}} x_0^4 $ and $b_i = 0$.

For the sake of readability of the manuscript we will summarize the most important results in the following and present all details in Appendix \ref{app:MZ}.

\begin{itemize}
	\item When using a linear projection operator, i.e., $h_1(x) = x$, we can derive the Mori-Zwanzig GLE (\ref{eq:mz_gle}) and find for the memory kernel
	\begin{align}\label{eq:Kn_smallb}
	K_\text{MZ}(t) &= K_\text{NL}(t) + K_\text{n}(t)  \nonumber\\
	K_\text{n}(t) &= \frac{b_{\text{e}}^2}{\kB  T} \langle \tilde{x}_0 (t) \tilde{x}_0 (0) \rangle_0 + \mathcal{O}(b_{\text{e}}^4)
	\end{align}
	where $K_\text{NL}(t)$ is the non-linear memory kernel presented in Eq.~(\ref{eq:MemoryAnalytic}), $\tilde{x}_0(t)=3\kB T x_0(t)/a_{\text{e}} - x_0(t)^3$ and $\langle...\rangle_0$ denotes an equilibrium average in the linear system with $b_{\text{e}}=0.$ The result resembles a similar finding of Zwanzig \cite{Zwanzig2001} in deterministic systems.
	\item Using the non-linear projection technique of Ref.~\cite{Vroylandt2022_nonlinear} the GLE is given by the NL GLE presented in Eq.~(\ref{eq:nl_gle}). This result might seem trivial, but the path to obtain it is very non-trivial and can be applied much more generally. It shows that the formalism presented in Ref.~\cite{Vroylandt2022_nonlinear} can indeed be applied in practical situations to derive non-linear GLEs and thus provides a firm basis for the NL GLE. Importantly, the fact that Eq.~(\ref{eq:nl_gle}) has no position-dependence in the memory kernel is very specific to the SNCL model and cannot be generally assumed.
\end{itemize}

The analytical calculations presented above will be accompanied by computer simulations of the stochastic differential equations (\ref{eq:eom}) as described in the next section.

\section{Molecular Dynamics and Coarse-grained simulations}
\label{sec:simulations}

 The equations of motion (\ref{eq:eom}) are equivalent to the well known Langevin dynamics. Consequently, we can use established tools to construct efficient integrators and perform molecular dynamics (MD) simulations \cite{gronbech2013simple}. This enables us to sample efficiently both the properties of the canonical ensemble defined by these equations for various parameter sets but also to calculate dynamical properties such as the mean squared displacement $\text{MSD}(t)$, the non-Gaussian parameter $\alpha(t)$ and the ISF $F_{\text{s}}(q,t),$ as introduced above. See Appendix \ref{app:md_simulations} for details on the computer simulations.

\subsection{Reconstruction of memory and thermal fluctuations}
\label{sec:reconstruction}

From the simulation results we will additionally extract further dynamical correlation functions to reconstruct the memory kernel $K(t)$ and properties of the thermal fluctuations $\eta(t)$.  In recent years, a plethora of methods have been suggested to calculate memory kernels from computer simulations and experiments \cite{Shin2010,Rotenberg2014_inverse,Lesnicki2016_mh,jung2017iterative,vroylandt2022likelihood} as recently reviewed \cite{klippenstein2021introducing}. Here, we will use one of the simplest methodologies based on the inverse Volterra technique suggested in Ref.~\cite{Shin2010}. For a general conservative force $F^\text{C}(t)$, corresponding to the first term on the right-hand side of the GLEs (\ref{eq:mz_gle}) and (\ref{eq:nl_gle}) we define $\tilde{F}(t) = m \frac{\diff v_0(t)}{\difft} - F^\text{C}(t) $ and can thus write
\begin{equation}
\langle \tilde{F}(t) v_0(0) \rangle = - \int_{0}^{t} \diff s K(t-s) \langle v_0(s) v_0(0) \rangle
\end{equation}
since the thermal fluctuations are orthogonal to $v_0(0)$ \cite{Zwanzig2001,Shin2010}. If we choose $F_\text{NL}^\text{C}(t)=-V_{\text{e}}^\prime(x_0(t))$, the above memory kernel will correspond to $K_\text{NL}(t)$, and if we choose $F_\text{MZ}^\text{C}(t) = -\tilde{a}_{\text{e}} x_0(t)$ we determine $K_\text{MZ}(t)$. Similar to Ref.~\cite{Shin2010} we then derive the Volterra equation of second kind and discretize the convolution integral to finally find the inversion algorithm
\begin{align}\label{eq:Kernel}
K(i  \Delta t) &= \frac{1}{\langle v_0^2 \rangle} \left\{ \frac{1}{m_0} \langle \tilde{F}(i \Delta t) F_0(0) \rangle - \right. \nonumber\\ &\left. \frac{\Delta t}{m_0} \sum_{j=0}^{i-1} w_j \langle {F_0}( (i-j)\Delta t) v_0(0) \rangle K(j  \Delta t)  \right\} 
\end{align}
with initial condition $K(0) = \frac{\langle \tilde{F}(0) F_0(0) \rangle}{m_0 \langle v_0^2 \rangle }$ and weight factor $w_j$ with $w_0 = 1/2$ and $w_j = 1$ for $j > 0$. Here, $\Delta t$ is the MD time step and all discretized time correlation functions are directly determined using the MD simulations.

It should be emphasized that the above algorithm is additive in the following sense: If one decomposes the external force $F_\text{NL}^\text{C}(t)$ into a linear and a non-linear contribution, $F_\text{NL}^\text{C}(t) = F^\text{C}_\text{MZ}(t) + F^\text{C}_{\text{n}}(t)$, we get $\tilde{F}_{\text{MZ}}(t) = \tilde{F}_{\text{NL}}(t) + F_{\text{n}}(t)$. Separately applying Eq.~(\ref{eq:Kernel}), one can then directly derive the corresponding decomposition for the kernels $K_{\text{MZ}}(t) = K_{\text{NL}}(t) + K_{\text{n}}(t)$. We will use this additivity in the present manuscript to study the non-linear contribution to the memory kernel, $K_{\text{n}}(t)$, and compare it to Eq.~(\ref{eq:Kn_smallb}). This additivity is not specific to the simplified SNCL model but holds generally (see discussion in Appendix~\ref{app:decomposition}). In fact in Refs.~\cite{Rotenberg2014_inverse,10.1063/5.0143707} a similar approach was used to study memory arising from the attractive and repulsive part of a Lennard-Jones potential.

Having extracted the memory kernel, we can immediately infer the thermal fluctuations by subtracting the dissipative force from $\tilde{F}(t)$ at each time step,
\begin{equation}\label{eq:eta}
\eta(t) = \tilde{F}(t) + \int_{0}^{t} \diff s K(t-s) v_0(s).
\end{equation}
This enables us to systematically analyze the properties of the thermal fluctuations.

\subsection{GLE simulations} 

With the reconstructed memory kernel $K(t)$ and the properties of the thermal fluctuations $\eta(t)$ we can perform numerical simulations by integrating the GLE. For this, we use the same algorithm as presented in Ref.~\cite{jung2017iterative}. The simulation algorithm uses a discretization of the convolution integral, as applied above for the inverse Volterra technique, to calculate the dissipative force. Correlated thermal fluctuations with a given correlation function $C_\eta(t)$ are created by the Fourier transform technique \cite{barrat2011portable} as described in detail in Appendix A of Ref.~\cite{jung2017iterative}. The only assumption for this technique is that the thermal fluctuations are Gaussian distributed. This assumption will, in fact, become questionable for non-linear coupling as will be discussed in Sec.~\ref{sec:anharmonic_coupling}.

\section{ Memory and Dynamics in the Linear SNCL model }
\label{sec:linear}

\begin{figure}
	\includegraphics[scale=0.98]{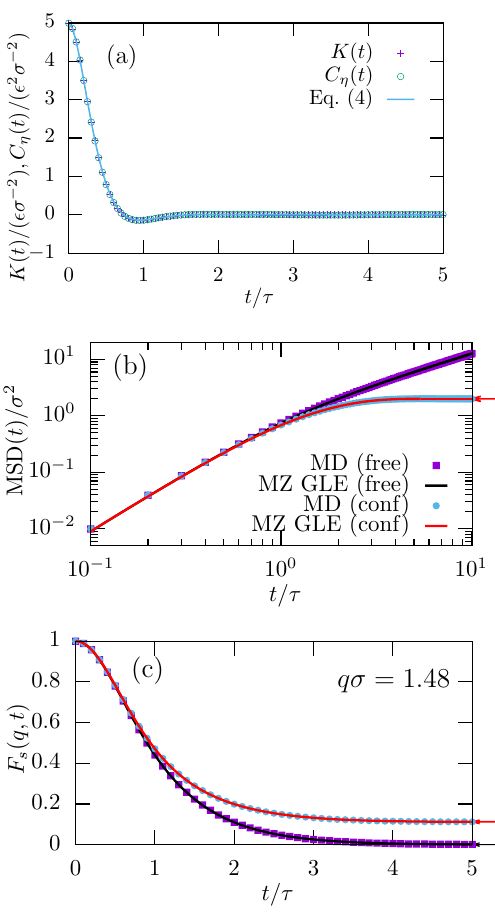}
	\caption{Dynamical correlation functions for a free ($a_{\text{e}}=0$) and a harmonically trapped ($a_{\text{e}}=1 \epsilon \sigma^{-2}$) particle with harmonic oscillator ($a_1=5\epsilon \sigma^{-2}$). (a) Reconstructed memory kernel $K(t)$ and noise correlation $C_\eta(t)$ as derived numerically using the algorithm in Eq.~(\ref{eq:Kernel}). The simulations are compared to the analytic results Eq.~(\ref{eq:MemoryAnalytic}). (b) Mean-square displacement (MSD) calculated in MD simulations and using the Mori-Zwanzig (MZ) GLE in Eq.~(\ref{eq:mz_gle}) with the memory shown in (a). The arrow indicates the limit $t\rightarrow \infty$ from Eq.~(\ref{eq:msd}). (c) Intermediate scattering function ($F_{\text{s}}(q,t)$) for the same trajectories as in (b). \gj{In this and the following figures, if the error bars are not visible they were smaller than the point size.} }
	\label{fig:linear}
\end{figure}

We set the stage and validate our methodology by investigating the dynamics of purely linear systems. We study the motion of a free and a harmonically trapped particle coupled to a single oscillator ($N=1$). We will use $N=1$ throughout this work since the results do not depend on the number of oscillators, apart from the analysis in Fig.~\ref{fig:largeN} where it is explicitly stated otherwise. We find that the reconstructed memory kernel for the motion of a tagged particle can indeed be quantitatively described by the analytical prediction Eq.~(\ref{eq:MemoryAnalytic}) (see Fig.~\ref{fig:linear}a). As expected, this result is independent of the external potential and the second fluctuation-dissipation theorem is fulfilled. 

Unlike the memory kernel, which describes only the dissipative motion, other dynamical quantities highlight the difference between the free and the trapped motion. The mean squared displacement $\text{MSD}(t)$, e.g., features a transition from short-time ballistic motion ($\text{MSD}(t) \propto t^2$) due to inertia to either long-time diffusion for free particles ($\text{MSD}(t) \propto t$) or trapped motion for confined particles ($\text{MSD}(t) = \text{const.}$), as shown in Fig.~\ref{fig:linear}b. This quantitative difference is also visible in the ISF, which -- depending on $q$ -- does not decay to zero in confinement (see Fig.~\ref{fig:linear}c). Importantly, the results from MD and GLE simulations are in perfect agreement indicating that the coarse graining procedure is \gj{very accurate}.

Having established these -- expected -- results for purely linear systems, we will now investigate how non-linearities affect the above observations.

\section{ Tagged particle in anharmonic trap ($b_{\text{e}} \neq 0$) }
\label{sec:anharmonic_trap}

\begin{figure}
	\includegraphics{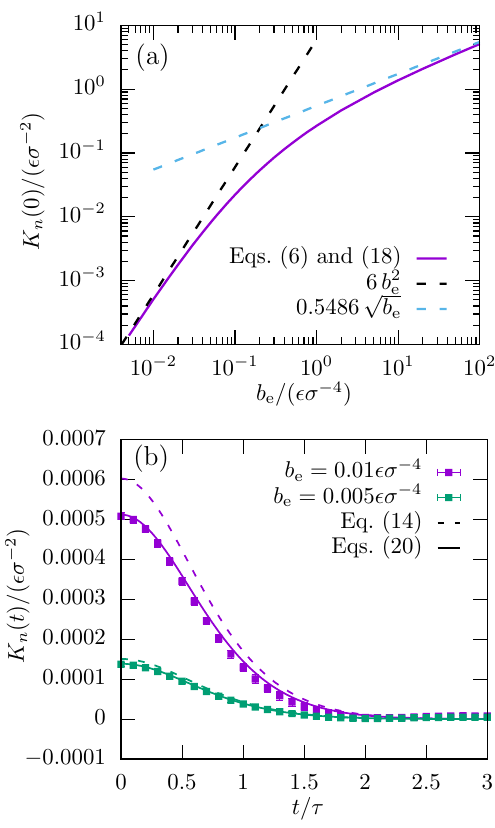}
	\caption{Non-linear contribution to the memory kernel, $K_{\text{n}}(t) = K_{\text{MZ}}(t) - K_{\text{NL}}(t)$, for different strengths of non-linearities of the external potential, $b_{\text{e}}$, while $a_{\text{e}}=1 \epsilon \sigma^{-2}$ (harmonic oscillator with $a_1=5\epsilon \sigma^{-2}$). (a) Instantaneous contribution to the memory kernel, $K_{\text{n}}(0)$, as extracted from numerical integration, Eq.~(\ref{eq:Kn_static}) (green line), Zwanzig scaling theory for $b \rightarrow 0$, Eq.~(\ref{eq:Kn_smallb}) with $\langle (\tilde{x}_0 (0))^2 \rangle_0 = 6$ (black line), and our scaling theory for $b \rightarrow \infty$ (orange line). (b) Full time-dependence of $K_{\text{n}}(t)$ for non-linear external potentials as extracted from MD simulations (symbols), Zwanzig scaling theory (dashed lines) and modified scaling theory (full lines). \gj{Magenta color here corresponds to $b_e=0.01 \epsilon \sigma^{-4}$ and green to $b_e=0.005 \epsilon \sigma^{-4}$.}
	}
	\label{fig:Kn}
\end{figure}

First, we study the impact of an anharmonic external potential. For this, we reconstruct both the non-linear GLE (\ref{eq:nl_gle}) with the memory kernel $K_\text{NL}(t)$ and the Mori-Zwanzig GLE (\ref{eq:mz_gle}) with $K_\text{MZ}(t).$ We are interested in the non-linear contribution to the memory kernel $K_{\text{n}}(t) = K_\text{MZ}(t) - K_\text{NL}(t),$ for which we have derived an analytical prediction in Eq.~(\ref{eq:Kn_smallb}) for small $b_{\text{e}}.$ We test the range of validity of this equation by investigating $K_{\text{n}}(0).$ For this special case we can derive an analytical expression,
\begin{equation}\label{eq:Kn_static}
K_{\text{n}}(0) = a_{\text{e}} - \tilde{a}_{\text{e}} + \frac{3 b_{\text{e}} \kB  T}{\tilde{a}_{\text{e}}}
\end{equation}
by using the relation 
\begin{equation}\label{eq:Kn_static2}
K_{\text{n}}(0) = \beta \langle F_{\text{n}}(t)^2  \rangle = - \left \langle \frac{\diff F_{\text{n}}(x_0)}{\diff x_0} \right\rangle = \langle V_{\text{e}}^{\prime\prime}(x_0) \rangle - \tilde{a}_{\text{e}}.
\end{equation}
Here, $F_{\text{n}}(t) = - V_{\text{e}}^{\prime}(x_0(t)) + \tilde{a}_{\text{e}} x_0(t) $ is the non-linear contribution to the external potential. The detailed {derivation} of Eqs.~(\ref{eq:Kn_static}) and (\ref{eq:Kn_static2}) is given in Appendix~\ref{app:instantaneous}.

Studying $K_{\text{n}}(0)$ we observe an intriguing transition from the expected scaling $K_{\text{n}}(0) \propto b_{\text{e}}^2$ for small $b_{\text{e}} \lesssim 0.01 \epsilon \sigma^{-4}$ to the large $b_{\text{e}}$ limit with scaling $K_{\text{n}}(0) \approx 0.5486 \sqrt{b_{\text{e}} \kB  T}$ (see Fig.~\ref{fig:Kn}a). The latter is obtained by inserting the limit of $\tilde{a}_{\text{e}}$ for $a_{\text{e}}=0$ (cf. Appendix \ref{app:atilde}) into Eq.~(\ref{eq:Kn_static}). This scaling is perfectly reproduced by the numerical solution shown as full line. Based on these results we investigate the time-dependence of $K_{\text{n}}(t)$ for small $b_{\text{e}} \leq 0.01 \epsilon \sigma^{-4}.$ We find that the theoretical prediction Eq.~(\ref{eq:Kn_smallb}) in this range indeed nicely describes the non-linear contribution to the memory kernel (compare dashed lines and data points in Fig.~\ref{fig:Kn}b). Interestingly, we observe that the main discrepancy between theory and simulation is already the approximate nature of the instantaneous contribution $K_{\text{n}}(0)$ in linear order of $b_{\text{e}}$. If we correct this value by using the exact prediction for $K_{\text{n}}(0)$ (Eq.~(\ref{eq:Kn_static})),
\begin{equation}\label{eq:corr_zwanzig}
	\gj{K_{\text{n}}(t) \approx \frac{a_{\text{e}} - \tilde{a}_{\text{e}} + 3 b_{\text{e}} \kB T / \tilde{a}_{\text{e}}}{\langle (\tilde{x}_0 (0))^2 \rangle_0} \langle \tilde{x}_0 (t) \tilde{x}_0 (0) \rangle_0}
\end{equation}
we find a significantly improved agreement (compare full lines and data points in Fig.~\ref{fig:Kn}b).

\begin{figure}
	\includegraphics{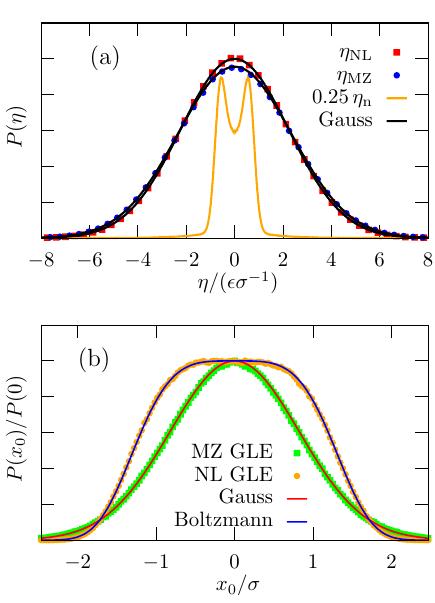}
	\caption{ Probability distributions of thermal fluctuations $\eta$ (a) and position $x_0$ (b) for a non-linear external potential, $a_{\text{e}}=0, b_{\text{e}}=1\epsilon \sigma^{-4}$. (a) Distribution of thermal fluctuations for the different GLEs presented in Sec.~\ref{sec:GLES}, as well as of the noise $\eta_{\text{n}}$ connected to the non-linear kernel $K_{\text{n}}(t) = K_{\text{MZ}}(t) - K_{\text{NL}}(t)$. \gj{(See definitions and explanations in Appendix \ref{app:decomposition}.)} $P(\eta_{\text{n}})$ is scaled with $0.25$. \gj{Probability distributions were calculated using histograms with high resolution of bins and long simulations to minimize statistical errors (not visible due to point size).} (b) Distribution of the position for the Mori-Zwanzig (MZ) and the non-linear (NL) GLE, Eqs.~(\ref{eq:mz_gle}) and (\ref{eq:nl_gle}), compared to the expected Boltzmann distribution $P(x_0) \propto e^{-\beta V_{\text{e}}(x_0)}$ and a corresponding Gaussian.     }
	\label{fig:distribution_external}
\end{figure}

Fig.~\ref{fig:Kn} highlights that one can generally reconstruct linear GLEs and understand how the non-linear forces in the original stochastic model are renormalized to give additional contributions to the memory kernel, $K_{\text{n}}(t),$ in the CG MZ GLE. The question is, which impact has this renormalization on the properties and dynamics of the CG GLE? We first find that the probability distribution of the thermal fluctuations $\eta(t)$ is not strongly affected by the renormalization (see Fig.~\ref{fig:distribution_external}a). While there are small deviations between the exact $\eta_\text{NL}$ and the MZ GLE $\eta_\text{MZ}$, caused by the additional contributions \gj{$\eta_{\text{n}}(t)= \eta_{\text{MZ}}(t) - \eta_{\text{NL}}(t)$}, the discrepancy is very small and both distributions are Gaussian. This is surprising because the pure renormalized fluctuations $\eta_{\text{n}}(t)$ are strongly non-Gaussian and actually feature a two-peak distribution (see Fig.~\ref{fig:distribution_external}a). \gj{The bimodal shape of $P(\eta_{\text{n}})$ can be rationalized by pointing out that $\eta_{\text{n}}(t)$ originates from the non-linear force $F^\text{C}_\text{NL}(x_0(t))$. Since this function has two extrema at which the occurrence probability of the force is maximum, a bimodal shape emerges and the positions of these two peaks depend on the ratio $(\tilde{a}_{\text{e}} - a_{\text{e}})/b_{\text{e}}$. }

Using these noise distributions as input for the GLE simulations, we study the resulting probability distribution of the position $P(x_0)$ in Fig.~\ref{fig:distribution_external}b and find severe qualitative differences between the exact NL GLE and the MZ GLE. The result can be easily rationalized: Both GLEs describe an equilibrium system, thus the distribution is expected to follow a Boltzmann distribution: $P(x) \propto e^{-\beta V(x)}$, where $V(x)$ is the static potential which is different between the NL GLE and the MZ GLE and describes the data perfectly (see Fig.~\ref{fig:distribution_external}b).

\begin{figure}
	\includegraphics{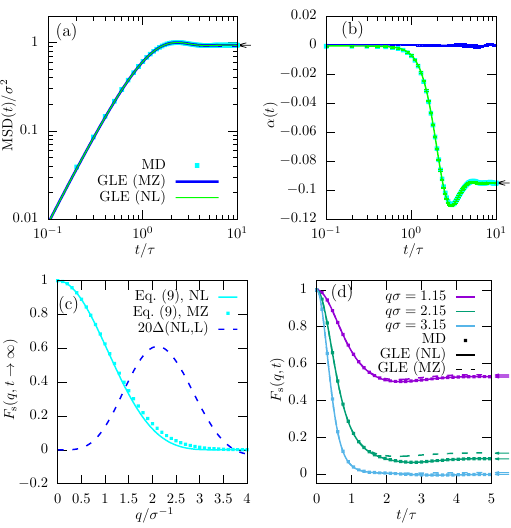}
	\caption{  Dynamical correlation functions for a non-linearly trapped particle ($a_\text{e} = 1 \epsilon \sigma^{-2}$, $b_\text{e}=1\epsilon \sigma^{-4}$) with harmonic oscillator. (a) Mean-squared displacement (MSD) as extracted from MD simulations, MZ and NL GLE. (b) Non-Gaussian parameter $\alpha(t)$. The arrow marks the limiting value for $t \rightarrow \infty$ from Eq.~(\ref{eq:alpha}). (c) Theoretical predictions for the long-time behavior of the incoherent intermediate scattering functions (ISF) for MZ and NL. The dashed line shows the magnified difference between the two curves. (d) Same data as in (a,b) for the ISF at three different wave numbers. The arrows correspond to the limiting values shown in (c). }
	\label{fig:dynamics_external}
\end{figure}

Despite these severe differences in the static distributions, the mean squared displacement $\text{MSD}(t)$ is in perfect agreement between the two types of GLEs (see Fig.~\ref{fig:dynamics_external}a). In fact, this agreement is enforced since the projection operator formalism is constructed based on the velocity autocorrelation function, see Appendix \ref{app:MZ}, which is trivially connected to the MSD ($\frac{\diff^2}{\difft^2} \text{MSD}(t) = -2 \langle v_0(t) v_0(0) \rangle$). Higher-order correlation functions are, however, not correctly reproduced. The non-Gaussian parameter $\alpha(t)$, defined in Sec.~\ref{sec:observables}, is always zero for the MZ GLE because the equations are linear and the thermal fluctuations used as input to the simulations are Gaussian. In contrast, however, $\alpha(t)$ is highly non-trivial for the non-linear simulations and not even vanishes in the limit $t\rightarrow \infty$ (see Fig.~\ref{fig:dynamics_external}b). 

We can further calculate experimentally accessible quantities such as the incoherent scattering function. From the long-time limit, which we compute numerically using Eq.~(\ref{eq:ISF_limit}), we already observe that there are systematic differences between MZ and NL GLE, which peak around $q = 2 \sigma^{-1}$ but can also become negative for larger values of $q$ (see Fig.~\ref{fig:dynamics_external}c). Comparing these theoretical predictions to the GLE results for the long-time limit of the ISF, we observe perfect agreement between theory and simulations (see Fig.~\ref{fig:dynamics_external}d).

Based on the SNCL model the above analysis clearly shows the implications and error sources when using a linearized GLE to describe non-linear dynamics. Before discussing these results in more detail in Sec.~\ref{sec:discussion}, we will now investigate the impact of anharmonic coupling between particle and oscillators.
	
\section{ Anharmonic Coupling  ($b_i \neq 0$) }
\label{sec:anharmonic_coupling}

\begin{figure}
	\includegraphics{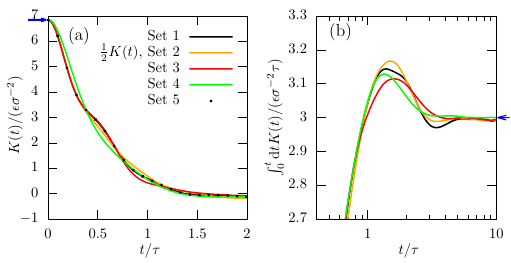}
	\caption{  Memory kernel $K(t)$ (a) and integrated memory kernel (b) for five different systems with non-linear coupling between tagged particle and oscillators. The left-hand-side arrow in (a) points to $K(0)$ calculated via Eq.~(\ref{eq:KvonNull}). The right-hand-side arrow in (b) indicates $ \int_{0}^{\infty} \text{\diff} t K(t) = \gamma_i$. The parameters for the different sets are summarized in Tab.~\ref{tab:parameters}.  }
	\label{fig:memory_coupling}
\end{figure}

\renewcommand{\arraystretch}{1.1}
\begin{table}
	\centering \begin{tabular}{c|c|c|c|c|c|c}

		 & $N$ & $a_{\text{e}}/(\epsilon \sigma^{-2})$ & $m_0/m$ &  $\kB T/\epsilon$ & $b_i/(\epsilon \sigma^{-4})$  \\
		\myrowcolour
	Set 1	& 1 &  2 & 1 & 1 & 11.42 \\ 
	Set 2	& 2 & 2  & 1  & 1 & 11.42\\ 
		\myrowcolour
	Set 3	& 1 & 0 &  1 & 1 & 11.42   \\ 
	Set 4	& 1 & 2 & 10 & 1 &  11.42 \\
		\myrowcolour
	Set 5	& 1 & 2 & 1 & 2 & 5.71 \\
	\end{tabular} 
	\caption{Summary of the five different systems used for the analysis in Fig.~\ref{fig:memory_coupling}. Other parameters were $b_{\text{e}} = 0, a_i=0, m_i = 1m, \gamma_i = 3m \tau^{-1} = 3 \epsilon \sigma^{-2} \tau$.}
	\label{tab:parameters}
\end{table} 
\renewcommand{\arraystretch}{1.0}

Different from the case of an anharmonic external potential, in which we could derive the NL GLE analytically, our analysis of anharmonic coupling will be mainly based on numerical simulations. Since the external potential will be linear in the present section, no difference exists between NL GLE and MZ GLE and we will therefore drop these subscripts in notation. 

First, we reconstruct the memory kernel $K(t)$ for different sets of parameters. The result highlights a fundamental difference between the two kinds of non-linearities studied within this work. While the anharmonic external potential did not influence the memory kernel if the oscillator was coupled harmonically, the situation is fundamentally different for anharmonic coupling. We find that the kernel $K(t)$ now depends on all parameters we have investigated: the strength of the (harmonic) external potential $a_{\text{e}}$, the mass of the tagged particle $m_0$, the temperature $\kB  T$ and the strength of the anharmonic coupling $b_i$. Further, the memory kernel is not additive anymore in the sense that it can be written as the sum of the individual oscillator contributions as in Eq.~(\ref{eq:MemoryAnalytic}), see Fig.~\ref{fig:memory_coupling}a. These results show that already such a simplified model can lead to complicated behavior of the memory kernel. There remain three relations for the memory kernel which do not depend on the parameters mentioned: The initial values of the kernel are additive and can be calculated exactly
\begin{align}\label{eq:KvonNull}
	K(0) = \sum_{i=1}^{N} a_i + 3b_i \kB T/\tilde{a}_i
\end{align}    
with $\tilde{a}_i =  \kB T/\langle (\Delta x_i)^2 \rangle$ (cf. Appendix \ref{app:instantaneous}).
An other invariance, which we can identify both using theory and simulations, is that the memory kernel does not change if $b \kB  T = \text{const.}$ for all $b_{\text{e}}$ and $b_i$ (see Fig.~\ref{fig:memory_coupling}a and Appendix \ref{app:invariants}). Furthermore, we find empirically for the long-time friction coefficient $\int_{0}^{\infty} \difft K(t) = \sum_{i=1}^{N} \gamma_i$ as in the case of a fully harmonic system. 

An interesting dependence of a memory kernel on an external potential has been discussed in Ref.~\cite{Daldrop2017_external} for the dynamics of solutes in water. There, however, the long-time friction coefficient $\gamma$, calculated as the total integral over the memory kernel, was found to increase with the external trapping potential. This seems to be in contrast to our finding in Fig.~\ref{fig:memory_coupling}b. The authors of Ref.~\cite{Daldrop2017_external} explain their observation by an increasing thickness of the hydration shell of the solute. This explanation can be well captured within our toy model: A larger hydration shell implies that the solute is in contact to an increasing number of solvent molecules. Translated into our model this means an increase in the number of coupled oscillators $N$ and thus larger friction $ \gamma = \sum_{i=1}^{N} \gamma_i$.

\begin{figure}
	\includegraphics{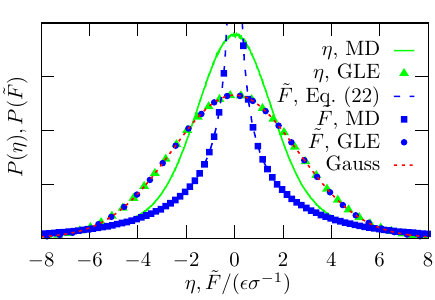}
	\caption{Probability distribution of thermal fluctuations $\eta$ and non-conservative forces $\tilde{F}$. MD simulations are compared to GLE. The dashed blue line shows the theoretical prediction for $\tilde{F}$ (Eq.~(\ref{eq:PFtilde})). (Parameters: $a_{\text{e}}=1\epsilon \sigma^{-2}$, $b_{\text{e}}=0$, $a_1=0$, $b_1=11.42\epsilon \sigma^{-4}$)   }
	\label{fig:distribution_coupling}
\end{figure}

Similar to the previous section we also investigate the probability distribution of the thermal fluctuations $P(\eta)$. We find that -- due to the anharmonic coupling -- $P(\eta)$, \gj{as derived from MD simulations using Eq.~(\ref{eq:eta}) (see green full line in Fig.~\ref{fig:distribution_coupling})}, deviates from a Gaussian distribution. Such non-Gaussian fluctuations cannot, within the current methodology, be included into a CG GLE model. We will therefore approximate it by a Gaussian fulfilling the second fluctuation-dissipation theorem (\gj{see $\eta$, GLE, in} Fig.~\ref{fig:distribution_coupling}).

We also observe that the probability distribution of the non-conservative forces $\tilde{F} = m \frac{\diff v_0(t)}{\difft} + V_{\text{e}}^\prime(x_0(t))$ determined in MD simulations deviates even stronger from a Gaussian distribution.  In fact, it even diverges for $ P(\tilde{F} \rightarrow 0)$ if $a_1=0$. This result can be derived analytically for $N=1$ using the fact that the non-conservative force can be written solely in terms of the relative distance between particle and oscillator, $\tilde{F} = \tilde{F}(\Delta x = x_0-x_1) = - V_1^\prime(\Delta x)$:
\begin{align}
P(\tilde{F}(\Delta x)) &= P(\Delta x) \left| \frac{d \Delta x}{d \tilde{F}} \right| \nonumber\\
& \propto \exp \left[ {-\beta V_1(\Delta x)} \right]  \left| \tilde{F} \right|^{-\frac{2}{3}} \nonumber \\
& \propto \exp \left[ {- { \frac{1}{4} \beta{\left |\tilde{F} \right |^\frac{4}{3}}{ b_1^{-\frac{1}{3}}}}} \right] \left| \tilde{F} \right|^{-\frac{2}{3}}
\end{align}
Importantly, however, this distribution can be integrated,
\begin{align}\label{eq:PFtilde}
\int_{-\infty}^{\infty} \diff \tilde{F} \exp \left[ {- { \frac{1}{4} \beta{\left |\tilde{F} \right |^\frac{4}{3}}{ b_1^{-\frac{1}{3}}}}} \right] \left| \tilde{F} \right|^{-\frac{2}{3}} = \frac{3 \, \Gamma(\frac{1}{4}) \, b_1^{\frac{1}{12}}}{\sqrt{2} \, \beta^{\frac{1}{4}}}   
\end{align}

\begin{figure}
	\includegraphics{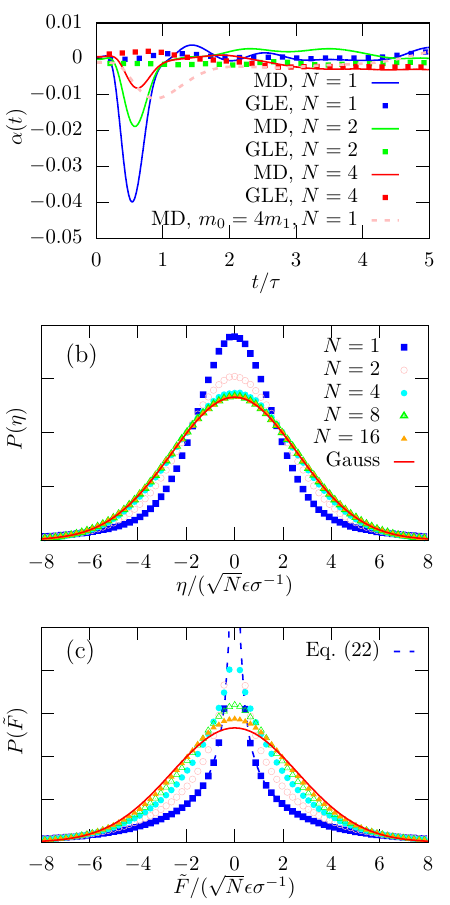}
	\caption{ Non-Gaussian dynamics $\alpha(t)$ (a) as well as probability distributions of thermal fluctuations $\eta$ (b) and non-conservative force $\tilde{F}$ (c) of the tagged particle with increasing number of oscillators $N$ and particle mass $m_0=Nm_1=0.2Nm$, other parameters as in Fig.~\ref{fig:distribution_coupling}. (a) compares the results from MD simulations to the GLE. As reference the dashed line shows also the MD result for a single oscillator with $m_0=4m_1.$  }
		\label{fig:largeN}
\end{figure}

We find perfect agreement between theory and simulations in Fig.~\ref{fig:distribution_coupling}. Contrarily, this force distribution is Gaussian for the CG GLE, as a consequence of the Gaussian velocity distribution and the Gaussian thermal forces. Our results thus clearly indicate that some central dynamical quantities might not be correctly reproduced in standard CG GLEs. This is supported by the non-Gaussian parameter $\alpha(t)$ in Fig.~\ref{fig:largeN}a (compare blue line and data points) which features a non-trivial time-dependence in the MD simulations. An additional analysis of the dynamics in such anharmonic systems is given in Appendix \ref{app:anharmonic_dynamics}.

To analyze whether these results can be generalized, we now study $N$ oscillators coupled to the tagged particle. The question is whether the dynamics of MD and GLE simulations become increasingly similar if we investigate more realistic scenarios in which a tracer interacts with many surrounding solvent particles at the same time. The results are shown in Fig.~\ref{fig:largeN}. We observe that the signal in the non-Gaussian parameter is significantly reduced and can thus be better approximated by the GLE with increasing $N$. Most importantly, we can rationalize this result by investigating the distributions of thermal and non-conservative forces. Both clearly converge towards a Gaussian distribution for increasing $N$, which is in agreement with the central limit theorem (see Fig.~\ref{fig:largeN}b,c). The non-conservative forces $\tilde{F}$ also loose their divergence at $\tilde{F}=0$ in this process and approach to a Gaussian distribution. 

The above results highlight that there are regimes in which the GLE can be used as an appropriate approximation of the underlying MD dynamics, in particular when investigating heavy molecules (such as larger polymers or colloids) which interact with many surrounding particles simultaneously.

\section{Discussion: Impact of non-linearities on GLE models}
\label{sec:discussion}

As mentioned in the Introduction, the main purpose of this work is to use a minimal but non-trivial toy system to better understand the consequences of dynamic coarse graining based on parameterization of the GLE. Thus we will discuss and generalize some of our findings.

The main conclusions of Section \ref{sec:anharmonic_trap} are that the correct handling of the conservative force is key in our model to properly describe static and dynamic properties of the system and that the purely linear MZ GLE fails to describe important properties such as non-Gaussian distributions for positions and displacements. We believe that this is fully generalizable to realistic soft matter systems. If the thermal force fulfills the second fluctuation-dissipation theorem, the GLE describes an equilibrium system, thus the static properties are fully (and uniquely) described by the conservative forces. Very recently, the correct derivation of GLEs with consistent potential of mean force has been discussed intensively in the literature \cite{Glatzel_2021,Vroylandt2022_nonlinear,Ayaz2022_GLENonLinear,Vroylandt_2022}. We have used one of the suggested extended formalisms \cite{Vroylandt2022_nonlinear} to derive a non-linear GLE and we believe the present manuscript further highlights the importance of such approaches, even if it comes with the cost of using \gj{position-dependent memory kernels} \cite{stella2014generalized,ness2017nonequilibrium,karniadakis2017_melt,jung2018_GLD,Vroylandt2022_nonlinear,Vroylandt_2022}. \gj{Potential techniques to extract such kernels have been proposed in Refs.~\cite{jung2017iterative,jung2018_GLD} and \cite{Vroylandt2022_nonlinear}.}

Section \ref{sec:anharmonic_coupling} then handles the problem of complex coupling to the bath particles. Despite the simplicity of the model we are able to analyze phenomena observed for realistic soft matter systems such as the memory kernel depending on external potentials \cite{Daldrop2017_external} and the non-Gaussian distribution of thermal fluctuations \cite{Shin2010}. The results clearly highlight that while the GLE is able to describe two-point autocorrelation functions as, e.g., the MSD, the correct treatment of higher-order correlation functions and complex dynamical properties is not implied. We assume that also this conclusion is generalizable: Systems, in which non-Gaussian displacements are not essential such as (dilute) colloidal or polymeric suspensions \cite{Shin2010,karniadakis2017_melt,jung2018_GLD,deichmann2018_polymer,klippenstein2021_ccc}, can be modeled using the currently existing methodology based on Gaussian thermal noise. There is, however, a plethora of soft matter and biological systems in which non-Gaussian dynamics are essential, including \gj{deeply supercooled liquids} \cite{Flenner2005_nonGaussian}, polymer melts \cite{bras2014_nonGaussian}, tracers in solutions of filaments \cite{wang2012_nonGaussian} and hydrogels \cite{cherstvy2019_nonGaussian}. Such systems should be treated with care in dynamic coarse graining and it should be analyzed specifically whether the microscopic dynamics can indeed be properly represented by the reconstructed GLE.

\section{Conclusion}
\label{sec:conclusion}

We have studied a simplified model system for the motion of tracer particles in complex environments relevant for biological and chemical soft matter physics. Using analytical theory, projection operator formalisms and computer simulations, we have investigated the consequences of emerging non-linearities in the equations of motion and how they impact reconstructed GLEs. As discussed in detail in the previous Section, our results highlight the importance of properly incorporating non-linearities into the GLE, e.g., using the recently introduced non-linear projection formalism \cite{Vroylandt2022_nonlinear}. We further have emphasized the importance of measuring higher-order correlation functions of dynamic CG models, which are much more difficult to reproduce than the commonly used two-point correlation functions such as mean squared displacement or velocity autocorrelation.

Our work also highlights open problems in dynamic CG modeling. Reconstructing non-linear potentials of mean force for strongly interacting systems is an active area of research, which should be directly applicable to coarse graining based on the GLE \cite{brini2019_cg,klippenstein2021introducing}. However, it is not obvious how to construct GLEs which properly reproduce non-trivial dynamical descriptors such as non-zero non-Gaussian parameters induced by the omnipresent non-linear interactions between the tagged particle and its surrounding, as analyzed in Sec.~\ref{sec:anharmonic_coupling}. Likely, this will require using more complex and realistic protocols for the thermal force, such as non-Gaussian, time-correlated random numbers. Recently, a promising approach to produce such random numbers using recurrent neural networks has been proposed \cite{zhu2022LSTM} and it will be interesting to investigate the potential of such techniques.

The situation becomes even more complex in non-equilibrium systems in which the separation between potentials of mean force, dissipative interactions and thermal forces becomes much less clear. Due to obvious applications in soft matter physics, such systems have recently seen an increasing attention also from the side of dynamic coarse graining using the GLE \cite{Jung_2022,maes2013fluctuation,meyer2017non,widder2022noise,vroylandt2022likelihood}. Performing similar analysis such as the one presented above to understand the impact of non-linearities in non-equilibrium systems is therefore also an exciting path of research.

\appendix

\section{Molecular dynamics (MD) simulations}
\label{app:md_simulations}

We have performed MD simulations to numerically solve the stochastic differential equations (\ref{eq:eom}). For this purpose,
based on Ref.~\cite{gronbech2013simple}, we derive a simple Langevin integrator:
\begin{align}\label{eq:Farago}
x_i(t) &= x_i(t-\Delta t) + f_i \Delta t v_i(t-\Delta t) + \frac{f_i \Delta t^2 F_i(t-\Delta t)}{2 m_i} \nonumber \\
& + \frac{f_i \Delta t W_i(t-\Delta t)}{2 m_i}   \\
v_i(t) &= v_i(t-\Delta t) + \frac{\Delta t}{2 m_i} (F_i(t-\Delta t) + F_i(t)) \nonumber \\
& - \frac{\gamma_i}{m_i} (x_i(t) - x_i(t-\Delta t)) + \frac{W_i(t)}{m_i}
\end{align}
with integration constant $f_i = \bigl( 1 + \frac{\gamma_i \Delta t}{2 m_i} \bigr)^{-1}$, the conservative forces $F_i(t) = -V_i^\prime(\Delta x_i(t))$ and Gaussian white noise $W_i(t)$. These equations are used for the oscillators ($i>0$). If $\gamma=0$ and thus $W(t)=0$ and $f=1$, they reduce to the standard velocity-Verlet algorithmus \cite{gronbech2013simple}, which is applied to the tagged particle (with $F_0(t) = -V_e^\prime(x_0(t)) -\sum_i V_i^\prime(\Delta x_i(t))$).

We choose $\Delta t = 0.005 \tau$ throughout this work. For $W_i$, normally distributed random numbers with variance $2 \gamma_i \kB T$ were used.

\section{Decomposition of memory kernels}
\label{app:decomposition}

Some simple rules for the computational handling of memory kernels and thermal fluctuations can be derived.

We start by adopting an alternative view regarding the external potential. We can look at it as the potential between an oscillator with infinite mass -- and therefore immobile -- and the particle, with zero friction coefficient.  

In this way, we can consider even the external potential as internal and calculate its kernel via Eq.~(\ref{eq:Kernel}) by inserting there $\tilde{F}(t) = V_{\text{e}}^{\prime}(x(t))$. We will refer to the resulting ''external'' memory kernel as $K_{\text{e}}(t)$. If the external potential is, e.g., purely harmonic with force constant $a_{\text{e}}$, we can deduce from Eq.~(\ref{eq:MemoryAnalytic}) that $K_{\text{e}}(t) = a_{\text{e}}$, i.e., the kernel calculated from an external linear force ($-a_{\text{e}} x_0$) equals the corresponding force constant.

Now, it also makes sense to insert $\tilde{F}(t) = F_0(t)$ in Eq.~(\ref{eq:Kernel}), thus obtaining a kernel that we call $K_{\text{tot}}(t)$.

\begin{figure}
	\hspace*{-0.8cm}	\includegraphics{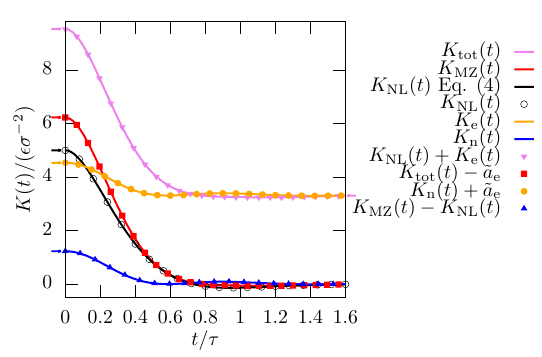}
	\caption{Several kernels, their sums and differences for the system $a_{\text{e}}=0$, $b_{\text{e}}=5\epsilon \sigma^{-4}$, $a_1=5\epsilon \sigma^{-2}$, $b_1=0$, as determined from MD simulations. The arrows on the left-hand side point to the theoretical values of the corresponding kernels, calculated via Eq.~(\ref{eq:instantaneous_papa}), the one on the right-hand side to $\tilde{a}_{\text{e}}$.   }
	\label{fig:decomposition}
\end{figure}

How are all these kernels connected to each other? Indeed, a simple relation can be derived: If a force can be decomposed into two components, $F_{\alpha} = F_{\beta} + F_{\gamma}$, then also the correlation function decomposes: $\langle F_{\alpha}(t)F_0(t) \rangle = \langle F_{\beta}(t)F_0(t) \rangle + \langle F_{\gamma}(t)F_0(t) \rangle$. If we insert this as $\langle \tilde{F}(t)F_0(t) \rangle$ into Eq.~(\ref{eq:Kernel}), we can see that the kernel decomposes the same way: $K_{\alpha}(t) = K_{\beta}(t) + K_{\gamma}(t)$. In Fig.~\ref{fig:decomposition} some results of this relation are shown: $K_{\text{tot}}(t) = K_{\text{NL}}(t) + K_{\text{e}}(t)$, $K_{\text{e}}(t) = K_{\text{n}}(t) + \tilde{a}_{\text{e}}$ (since $- V_{\text{e}}^{\prime}(x_0(t)) = F_{\text{n}}(t) - \tilde{a}_{\text{e}} x_0(t)$, cf. Sec.~\ref{sec:anharmonic_trap}) and $K_{\text{MZ}}(t) = K_{\text{NL}}(t) + K_{\text{n}}(t)$.

If a kernel can be decomposed as shown above, then the corresponding thermal fluctuation and its autocorrelation can be decomposed, too (cf. Eqs.~(\ref{eq:eta}) and (\ref{eq:FDTAnalytic})), e.g., $\eta_{\text{MZ}}(t) = \eta_{\text{NL}}(t) + \eta_{\text{n}}(t)$ and $C_{\eta_{\text{MZ}}}(t) = C_{\eta_{\text{NL}}}(t) + C_{\eta_{\text{n}}}(t)$.

Caveat: The rules of decomposition introduced above are valid for the analysis of a given system with given parameters. They do not imply that, e.g., the kernel of an ensemble with two oscillators is the sum of the kernels of ensembles with the single ones. The latter is not true for anharmonic oscillators (cf. Fig.~\ref{fig:memory_coupling}).

\section{Extended derivations of analytical results}

In the main part of the manuscript we have derived several analytical results. Here we provide further details for their derivation.

\subsection{NL GLE for $b_i = 0$}
\label{app:analytic_bo0}

The equation of motion can be rewritten as a first-order differential equation, 
\begin{align}
\frac{\diff}{\difft} \left( \begin{array}{c}
x_0(t) \\
m_0 v_0(t)  \\
\vdots \\
x_N(t) \\
m_Nv_N(t)
\end{array} \right)&=
\left( \begin{array}{c}
0  \\
V_\text{e}'(x_0) \\
\vdots \\
0 \\
0
\end{array} \right) 
\\ &+ \nonumber
\left( \begin{array}{cccc}
0  & 1 & 0 & 0 \\
0 & 0 & \sum_i a_i & 0 \\
0 &0&0&1 \\
-a_i & 0 & 0 & 0
\end{array} \right)
\left( \begin{array}{c}
x_0(t) \\
v_0(t)  \\
\vdots \\
x_N(t) \\
v_N(t)
\end{array} \right)\\
&	+  \left( \begin{array}{ccccc}
0  & 0 & \cdots & 0 & 0 \\
0 & 0 & \cdots & 0 & 0 \\
\vdots & \vdots &  & \vdots & \vdots \\
0 &0& \cdots&0&0 \\
0 & 0& \cdots & 0 & \gamma_N
\end{array} \right)
\left( \begin{array}{c}
x_0(t) \\
v_0(t)  \\
\vdots \\
x_N(t) \\
v_N(t)
\end{array} \right) \nonumber\\&+\left( \begin{array}{ccccc}
0  & 0 & \cdots & 0 & 0 \\
0 & 0 & \cdots & 0 & 0 \\
\vdots & \vdots &  & \vdots & \vdots \\
0 &0& \cdots&0&0 \\
0 & 0& \cdots & 0 & \sqrt{2 \gamma_N \kB  T}
\end{array} \right)
\left( \begin{array}{c}
0 \\
0  \\
\vdots \\
0 \\
W_N(t)
\end{array} \right) \nonumber
\end{align}
with white noise $  \langle W_i(t)W_k(0) \rangle = \delta(t) \delta_{ik} $.

This equation is particularly interesting, since very similar equations of motion have been proposed and actively applied as extended Markov models where the oscillators represent auxiliary variables which ``mimic'' the memory of an otherwise non-Markovian dynamic system \cite{Ceriotti2010,karniadakis2017_melt,klippenstein2021introducing,ayaz2021non} . The detailed analysis presented in the main text is thus also relevant in terms of exploring the behavior of such auxiliary variable systems.

To derive the NL GLE we are only interested in the dynamics of $ x_i(t) $ since only this variable couples to the reference particle. After some algebra we find
\begin{align}\label{eq:solve_qk}
x_i(t) &= x_0(t)   - \frac{1}{m_i a_i} \int_0^t \diff s K_i(t-s)  v(s)\\ & +  \int_0^t \diff s \frac{\sqrt{2 \gamma_i \kB  T}}{\omega_i^l m_i} e^{-\delta_i (t-s)} \sin(\omega_i^l (t-s)) W_k(s) \nonumber
\end{align}
with
\begin{equation}\label{eq:memory}
K_i(t) =  a_i e^{-\delta_i t} \left[  \cos(\omega_i^l t) + \frac{\delta_i}{\omega_i^l}  \sin(\omega_i^l t)  \right]
\end{equation}
where $ \delta_i = \frac{\gamma_i}{2 m_i} $, $\omega_i = \sqrt{a_i/m_i}$ and $ \omega_i^l = \sqrt{\omega_i^2 - \delta_i^2 } $. Here, we have assumed that the oscillators are underdamped (i.e., $ \omega_i > \delta_i $ ). In the critical case, i.e., if $ \omega_i = \delta_i $ and thus $ \omega_i^l = 0 $, Taylor expansion of the sinus term results in
\begin{equation}\label{eq:memory1}
K_i(t) = a_i e^{-\delta_i t} \left[ 1 + \delta_i t \right]
\end{equation}
In the overdamped case, i.e., if $ \omega_i < \delta_i $, complex algebra using $ \bar{\omega}_i^l = \sqrt{ \delta_i^2 - \omega_i^2 } $ gives
\begin{equation}\label{eq:memory2}
K_i(t) = a_ie^{-\delta_k t} \left[ e^{-\bar{\omega}_i^{l} t} \left( \frac{1}{2} - \frac{\delta_i}{2 \bar{\omega}_i^{l}} \right) + e^{\bar{\omega}_i^{l} t} \left( \frac{1}{2} + \frac{\delta_i}{2 \bar{\omega}_i^{l}} \right) \right]
\end{equation}

Inserting Eq.~(\ref{eq:solve_qk}) into the equations of motion for the reference particle we find
\begin{equation}\label{eq:final_eom}
m_0\frac{\diff}{\difft} v_0(t) = - V_\text{e}'(q) - \int_0^t \diff s K(t-s)  v(s) + \eta(t)\\
\end{equation}
where $ K(t) = \sum_i K_i(t) $, and the noise is given by the autocorrelation function
\begin{align}
\langle \eta(t) \eta(t^\prime) \rangle = {\kB  T} \sum_k  K_k(t-t^\prime) = {\kB  T}   K(t-t^\prime)
\end{align}
Thus the fluctuation-dissipation theorem is fulfilled (for large $ t $).

\subsection{Effective harmonic strength, $\tilde{a}$}
\label{app:atilde}

In the following, the effective harmonic strength can be related to the external potential. Then $a,b,x,\tilde{a}$ mean $a_{\text{e}}, b_{\text{e}},x_0,\tilde{a}_{\text{e}}$. But it can also be related to the coupling potential between particle and oscillator. Then $a,b,x,\tilde{a}$ stand for $a_i, b_i,x_i,\tilde{a}_i$.

By definition of $\tilde{a}$, the following equipartition theorem must hold:
\begin{align}\label{eq:atilde1}
	a \langle x^2 \rangle + b \langle x^4 \rangle =  \kB T = \tilde{a} \langle x^2 \rangle
\end{align}
Introducing the dimensionless parameter $\theta = \frac{\langle x^4 \rangle}{\langle x^2 \rangle ^2}$ gives
\begin{align}\label{eq:atilde2}
	\tilde{a} = a + b \theta \langle x^2 \rangle = a + b \theta  \kB T/\tilde{a}
\end{align}
Solving this quadratic equation results in
\begin{align}\label{eq:atilde3}
	\tilde{a} = \frac{a}{2} + \sqrt{\frac{a^2}{4} + \theta b \kB T}
\end{align}
Eq.~(\ref{eq:atilde3}) is exact, without approximation. All we don't know is in $\theta$.
For a purely harmonic potential ($b=0, \theta=3$), $\tilde{a} = a$. For a purely anharmonic potential ($a=0, \theta = \frac{ \Gamma(0.25)^4}{{8} \pi^2} \approx 2.18844$), $\tilde{a} = \sqrt{\theta b \kB T} \approx 1.479 \sqrt{b  \kB T}$.

Series expansion of Eq.~(\ref{eq:atilde3}) to first order in $b$ using $\theta = 3 + \mathcal{O}(b^1)$ gives (cf. Eq.~(\ref{eq:pmflinear})):
\begin{align}\label{eq:atilde4}
	\tilde{a} = a + \frac{3b \kB T}{a} + \mathcal{O}(b^2)
\end{align}

\subsection{Non-linear projection operator formalism}
\label{app:MZ}

\gj{The non-linear projection formalism in Ref.~\cite{Vroylandt2022_nonlinear} starts} by defining functions which select the positions and velocities of the tagged particle,
\begin{align}\label{key}
&\mathcal{O}_x(\bm{X})= x_0 \\
&\mathcal{O}_v(\bm{X})=\mathcal{K}\mathcal{O}_x(\bm{X}) = v_0.
\end{align}

To apply the projection operator formalism, we \gj{then introduce} the operator
\begin{equation}\label{eq:projector}
\mathcal{P}_\mathcal{E} \varphi = \sum_{k\in \mathcal{I}} \left( \sum_{k^\prime\in \mathcal{I}} G^{-1}_{k,k^\prime} \langle E_{k^\prime}, \varphi \rangle \right) E_k
\end{equation}
for any observable $\varphi: \mathbb{R}^{2(N+1)} \rightarrow \mathbb{R} $. Here, $\mathcal{E} = \{ e_k,k\in \mathcal{I}\}$ is the set of functions $e_k: \mathbb{R}^{2} \rightarrow \mathbb{R}^2 $ on which we project the dynamics, $\mathcal{I}$ is a set of indices, $E_k = e_k(\mathcal{O}_x(\mathbf{X}),\mathcal{O}_v(\mathbf{X}))$ and the Gram matrix of the basis is given by $G_{k,k^\prime} = \langle E_k, E_{k^\prime} \rangle$. \gj{This form of the projection operator is very similar to the linear Mori projector \cite{Mori1965,Zwanzig2001}, but it leaves open the possibility to introduce non-linear functions $E_k.$}

Following Ref.~\cite{Vroylandt2022_nonlinear} we choose \gj{the basis functions}
\begin{align}
\mathcal{E} = &\left\{  a: \begin{pmatrix} o_1 \\ o_2 \end{pmatrix} \mapsto \begin{pmatrix} o_1 \\ 0 \end{pmatrix} \right\} \cup \left\{  b_k: \begin{pmatrix} o_1 \\ o_2 \end{pmatrix} \mapsto \begin{pmatrix} 0 \\ h_k(o_2) \end{pmatrix} \right\}_{k\in \mathcal{J}} \nonumber \\ & \cup \left\{  c_k: \begin{pmatrix} o_1 \\ o_2 \end{pmatrix} \mapsto \begin{pmatrix} 0 \\ h^\prime_k(o_1)o_2 \end{pmatrix} \right\}_{k\in \mathcal{J}}
\end{align}
with the family of functions $\{h_k, k\in \mathcal{J}\}$ from $\mathbb{R} \rightarrow \mathbb{R}$ and index set $\mathcal{J}$. We will specify these functions depending on whether we use a linear projection ($h_1(x) = x_0$), \gj{which would correspond to the original Mori-Zwanzig formalism \cite{Zwanzig2001}} or a non-linear projection ($h_1(x) = x_0$, $h_2(x) = x_0^3$), \gj{allowing us to introduce a non-linear potential of mean force into the formalism.}

In the following, we present detailed derivations for the results presented in Sec.~\ref{sec:MZ}. The derivation follows the line of Ref.~\cite{Vroylandt2022_nonlinear} and also adapts all definitions and nomenclature of this reference. We will present separate results using a linear projection ($h_1(x) = x_0$) or a non-linear projection ($h_1(x) = x_0$, $h_2(x) = x_0^3$).

\subsubsection{Mean force, $f_b(x_0)$}

According to Ref.~\cite{Vroylandt2022_nonlinear} the mean force is defined as 
\begin{equation}
f_b(x_0)=\sum_{k\in \mathcal{J}} \left(  \sum_{k^\prime \in \mathcal{J}}  (G_b^{-1})_{k,k^\prime} \langle B_k^\prime, \mathcal{K} \mathcal{O}_2 \rangle \right ) h_k.
\end{equation}

 \noindent\textbf{Linear projection:}
If we choose $\mathcal{I}={1}$ and $h_1(x) = x_0$ we find
\begin{equation}
G_b^{-1} = \langle x_0^2 \rangle ^{-1}.
\end{equation}
We can immediately evaluate $\mathcal{K} \mathcal{O}_v = \mathcal{K} v_0 = -V_{\text{e}}^\prime(x_0) + \sum_{i>0} a_{\text{e}} (x_i(t)-x_0(t))$. From this we determine 
\begin{equation}
\langle B_1, \mathcal{K} \mathcal{O}_v \rangle = -a_{\text{e}} \langle  x_0^2 \rangle - b_{\text{e}} \langle  x_0^4\rangle.
\end{equation}
The final result for the mean force is thus  
\begin{equation}\label{eq:pmflinearfull}
f_0(x_0)= (-a_{\text{e}} -b_{\text{e}} \langle  x_0^4\rangle/\langle  x_0^2\rangle) x_0 = - \tilde{a}_{\text{e}} x_0
\end{equation}   
with $\tilde{a}_{\text{e}} = a_{\text{e}} + b_{\text{e}} \langle  x_0^4\rangle/\langle  x_0^2\rangle $. To first order in $b_{\text{e}}$ we find (cf. Eq.~(\ref{eq:atilde4}))
\begin{equation}\label{eq:pmflinear}
f_0(x_0)= - (a_{\text{e}} + 3b_{\text{e}} \kB T/a_{\text{e}}) x_0 + \mathcal{O}(b^2).
\end{equation}

\noindent\textbf{Non-linear projection:}
If we choose $\mathcal{I}={1,2}$ with $h_1(x) = x_0$, $h_2(x) = x_0^3$, we find for the resulting mean force $f_b(x_0) = -V^\prime(x_0)$, as shown in the following. The conditional expectation $ \psi(x_0) $ is given by \cite{Vroylandt2022_nonlinear}
\begin{equation}\label{key}
\psi(x_0) = \mathbb{E}(\mathcal{K}_{ov} \varphi(\bm{X}) | \varphi(\bm{X}) = x_0)
\end{equation}
with 
\begin{equation}
\mathcal{K}_{ov} \varphi(\bm{X}) = - \nabla V(\bm{X}) \cdot \nabla \varphi(\bm{X}) + \frac{1}{\beta} \nabla \varphi(\bm{X}).
\end{equation}
Using $ \varphi(\bm{X}) = x_0 $, we find $\mathcal{K}_{ov} \varphi(\bm{X}) = - V^\prime(x_0)$ and thus $\psi(x_0) = - V_{\text{e}}^\prime(x_0) $.

Since $\psi(x_0)$ is therefore in the space spanned by ${h_k,k\in\mathcal{J}}$, as discussed in Sec. 2.4 of Ref.~\cite{Vroylandt2022_nonlinear}, we find $f_b(x_0) = \psi(x_0) = - V_{\text{e}}^\prime(x_0)$.

Importantly, this derivation immediately generalizes to any potential $V_{\text{e}}(x_0)$ up to order $ n $ in $x_0$, given that we choose $\mathcal{I} = {1,...,n-1}$ and $h_k(x) = x_0^k,$ even if $ n \rightarrow \infty. $ 

\subsubsection{Noise $\eta(t)$ and memory $K(t,x_0(t))$}

The noise $\eta(t)$ in the projection operator formalism is given by 
\begin{equation}\label{eq:defNoise}
\eta(t)= e^{t(1-\mathcal{P}_\epsilon)\mathcal{K}}(1-\mathcal{P}_\epsilon)\mathcal{K} v_0
\end{equation}
and follows the orthogonal dynamics
\begin{align}
\frac{\text{\diff}}{\text{\diff} t} \eta(t) = (1-\mathcal{P}_\mathcal{E}) \mathcal{K} \eta(t) \label{eq:orthDyn}\\ \eta(0) = (1-\mathcal{P}_\mathcal{E}) \mathcal{K} v_0.
\end{align}
It therefore includes combinations of the (generally) non-linear Fokker-Planck operator $\mathcal{K}$ and the (generally) non-linear projection operator $\mathcal{P}_\mathcal{E}$, as defined above.

We also calculate the memory kernel given by
\begin{equation}
K(t,x_0(t)) = \sum_{k\in \mathcal{J}} \left(  \sum_{k^\prime \in \mathcal{J}}  (G_c^{-1})_{k,k^\prime} \langle \mathcal{K} C_{k^\prime}, \eta(t) \rangle \right ) h^\prime_k(x_0(t)).
\end{equation}

In the following we first discuss the linear case, $V_{\text{e}}(x_0)=\frac{1}{2}a_{\text{e}}x_0^2$, using a linear projection scheme. We then generalize these results to non-linear potentials, similar in spirit to Zwanzig \cite{Zwanzig2001}. Finally we apply the non-linear projection technique to non-linear potentials.

\noindent\textbf{Linear projection, linear potential}
For the linear case the Fokker-Planck operator is explicitly given by
\begin{align}\label{eq:fplinear}
\mathcal{K}_0 &= v_0 \frac{\partial }{\partial x_0}  - a_{\text{e}} x_0 \frac{\partial }{\partial v_0} + \sum_{i>0} a_i (x_i-x_0) \frac{\partial }{\partial v_0}\nonumber\\ &+ \sum_{i>0}  v_i \frac{\partial }{\partial x_i} - \sum_{i>0}  a_i (x_i-x_0) \frac{\partial }{\partial v_i} - \frac{\gamma_i}{m_i} v_i \frac{\partial }{\partial v_i} \nonumber\\&+ T \sum_{i>0} \frac{\gamma_i}{m_i^2} \frac{\partial^2 }{\partial v_i^2} 
\end{align}
and we choose $\mathcal{I}={1}$ and $h_1(x) = x_0$. Importantly, it can be noticed that both the Fokker-Planck operator and the projection operator will map a linear function onto another linear function. If we therefore use a general ansatz,
\begin{equation}\label{eq:ansatz}
\eta_0(t) = \rho_0(t) x_0 + \sigma_0(t) v_0 + \sum_{i=1}^N\rho_i(t) x_i + \sum_{i=1}^N\sigma_i(t) v_i
\end{equation}
we know that the explicit evaluation of the right-hand side of Eq.~(\ref{eq:orthDyn}) will give a linear result which allows us to use this equation to explicitly derive differential equations for the time-dependent prefactors, $\rho_i(t)$ and $\sigma_i(t)$ \cite{Zwanzig2001,Jung_2022}. 

After some simple calculations, we find
\begin{align}
\mathcal{K}_0 \eta_0(t) = &\rho_0(t) v_0 - \sigma_0(t) \left( a_{\text{e}} x_0 + \sum_{i>0} a_i (x_i -x_0) \right) \nonumber\\&+ \sum_{i>0} \rho_i(t)  v_i - \sum_{i>0}\sigma_i(t) \left( a_i ( x_i - x_0) + \gamma_i v_i \right) \nonumber\\
\mathcal{P}_\mathcal{E}\mathcal{K}_0 \eta_0(t) =& \rho_0(t)  v_0 - \sigma_0(t) a_{\text{e}} x_0.
\end{align}
We can thus derive the differential equations
\begin{align}
\frac{\text{\diff}}{\text{\diff}t} \rho_0(t) &= \sum_{i>0}  a_i (\sigma_i(t) - m_i\sigma_0(t)) \quad &\rho_0(0) &= \sum_{i>0} a_i\\
\frac{\text{\diff}}{\text{\diff}t} \sigma_0 (t) &= 0 &\sigma_0(0) &= 0\\
\frac{\text{\diff}}{\text{\diff}t} \rho_i(t) &= -  a_i (\sigma_i(t) - m_i\sigma_0(t)) &\rho_i(0) &= - a_i \label{eq:dgl3}\\
\frac{\text{\diff}}{\text{\diff}t} \sigma_i(t) &= \rho_i(t) - \sigma_i(t) \frac{\gamma_i}{m_i} & \sigma_i(0) &= 0 \label{eq:dgl4}
\end{align}
We immediately conclude the trivial solution, $\sigma_0(t)=0$. This transforms Eqs.~(\ref{eq:dgl3}) and (\ref{eq:dgl4}) into the differential equations for a damped harmonic oscillator which can be easily solved (assuming here the underdamped case, $\delta_i = \gamma_i/2m_i < \omega_i$),
\begin{align}
\sigma_i(t)&= -\frac{a_i}{\tilde{\omega}_i}e^{-\delta_i t} \sin(\tilde{\omega}_i t) \\
\rho_i(t) &= -\frac{ \gamma_i  a_i}{2 \tilde{\omega}_i}e^{-\delta_i t} \sin(\tilde{\omega}_i t) -a_i e^{-\delta_i t} \cos(\tilde{\omega}_i t)
\end{align}
with $\omega_i = \sqrt{a_i/m_i}$ and $\tilde{\omega}_i = \sqrt{\omega^2_i - \delta_i^2}.$ We therefore find the final relation for the noise,
\begin{align}
\eta_0(t) = &- \sum_{i>0} \frac{a_i}{\tilde{\omega}_i}\sin(\tilde{\omega}_i t) e^{-\delta_i t} v_i\\
&-\sum_{i>0}^N \left(\frac{ \gamma_i  a_i}{2 \tilde{\omega}_i}e^{-\delta_i t} \sin(\tilde{\omega}_i t) +a_i e^{-\delta_i t} \cos(\tilde{\omega}_i t) \right)( x_i - x_0)
\end{align}
and thus the autocorrelation function,
\begin{equation}\label{eq:FDTProject}
\langle \eta_0(t) \eta_0(0) \rangle = \kB  T \sum_{i>0}^N a_ie^{-\delta_i t}\left(\frac{ \gamma_i  }{2 m_i \tilde{\omega}_i} \sin(\tilde{\omega}_i t) +  \cos(\tilde{\omega}_i t) \right)
\end{equation}
where we used $\langle ( x_i - x_0)( x_i - x_0)  \rangle = \frac{\kB  T}{a_i}$. This expression is clearly identical to the noise derived in Eq.~(\ref{eq:FDTAnalytic}).

To evaluate the memory kernel we realize that $G_c= \langle v_0^2\rangle$ and thus find
\begin{equation}
K_0(t,x_0(t)) = \langle v_0^2\rangle ^{-1} \langle (\eta_0(0) + \mathcal{P}_\mathcal{E} \mathcal{K}_0 v_0) \eta_0(t) \rangle.
\end{equation}
Since $\eta_0(t)$ is, per definition, orthogonal to $\mathcal{P}_\mathcal{E} \mathcal{K}_0 v_0$, we obtain the usual second fluctuation dissipation theorem
\begin{equation}\label{eq:MemoryProject}
\kB  T K_0(t,x_0(t)) =  \kB  T K_0(t) =\langle \eta_0(t) \eta_0(0) \rangle.
\end{equation}
The results of the projection scheme are therefore equivalent to the analytical model in Eq.~(\ref{eq:MemoryAnalytic}).

\noindent\textbf{Linear projection, non-linear potential:}
To handle the non-linear potential, it is necessary to separate the Fokker-Planck operator into a linear and a non-linear part,
\begin{equation}
\mathcal{K} = \mathcal{K}_0 + \mathcal{K}_1
\end{equation}
where $\mathcal{K}_0$ is as defined above in Eq.~(\ref{eq:fplinear}) and $\mathcal{K}_1 = -  b_{\text{e}} x_0^3 \frac{\partial }{\partial v_0} $. 

To calculate the noise, defined in Eq.~(\ref{eq:defNoise}), we separate $(1-\mathcal{P}_\epsilon)\mathcal{K} v_0$ into two contributions,
\begin{equation}
(1-\mathcal{P}_\epsilon)\mathcal{K} v_0 = (1-\mathcal{P}_\epsilon)(\mathcal{K}_0+\mathcal{K}_1) v_0
\end{equation}
where the second term can be easily evaluated to $(1-\mathcal{P}_\epsilon)\mathcal{K}_1v_0 = (\tilde{a}_{\text{e}} -a_{\text{e}})x_0-b_{\text{e}}x_0^3$ with $\omega_0$ as defined in Eq.~(\ref{eq:pmflinearfull}).

We can also expand the orthogonal time evolution operator,
\begin{align}
e^{t(1-\mathcal{P}_\epsilon)\mathcal{K}} &= e^{t(1-\mathcal{P}_\epsilon)\mathcal{K}_0} \\+& \int_0^\infty \text{\diff}s e^{(t-s)(1-\mathcal{P}_\epsilon)\mathcal{K}_0} (1-\mathcal{P}_\epsilon) \mathcal{K}_1 e^{s(1-\mathcal{P}_\epsilon)\mathcal{K}_0} + \mathcal{O}(b^2) \nonumber
\end{align}
Meaning that, to first order, the noise has three contributions,
\begin{align}
\eta(t) &= e^{t(1-\mathcal{P}_\epsilon)\mathcal{K}_0}(1-\mathcal{P}_\epsilon)\mathcal{K}_0 v_0 \\ +& e^{t(1-\mathcal{P}_\epsilon)\mathcal{K}_0}((\tilde{a}_{\text{e}} -a_{\text{e}})x_0-b_{\text{e}}x_0^3)  \nonumber\\
+ &\int_0^\infty \text{\diff}s e^{(t-s)(1-\mathcal{P}_\epsilon)\mathcal{K}_0} (1-\mathcal{P}_\epsilon) \mathcal{K}_1 e^{s(1-\mathcal{P}_\epsilon)\mathcal{K}_0}(1-\mathcal{P}_\epsilon)\mathcal{K}_0 v_0 \nonumber
\end{align}
The first term is identical to $\eta_0(t)$ that we have determined in the linear potential section above. The third term vanishes, 
\begin{equation}\label{eq:expand3}
(3rd) = \int_0^\infty \text{\diff}s e^{(t-s)(1-\mathcal{P}_\epsilon)\mathcal{K}_0} (1-\mathcal{P}_\epsilon) \mathcal{K}_1 \eta_0(s) = 0
\end{equation}
since $\eta_0(s)$ is independent of $v_0.$

The second term, to first order, is
\begin{equation}
\eta_{b}(t) = b_{\text{e}} e^{t(1-\mathcal{P}_\epsilon)\mathcal{K}_0} (3\kB  Tx_0/a_{\text{e}}-x_0^3) + \mathcal{O}(b^2)
\end{equation}
Clearly, we can write down the time derivative for $\eta_b(t)$,
\begin{eqnarray}
\frac{\diff}{\difft} \eta_{b}(t) = (1-\mathcal{P}_\epsilon)\mathcal{K}_0\eta_{b}(t)
\end{eqnarray}
As Zwanzig argues in Ref.~\cite{Zwanzig2001}, the term  $\mathcal{K}_0\eta_{b}(t)$ lies in the orthogonal subspace and $\mathcal{P}_\epsilon\mathcal{K}_0\eta_{b}(t)=0$. Therefore, the time evolution of $\eta_b(t)$ is actually given by the normal dynamics in the linear system with $b_{\text{e}}=0$, and we find
\begin{equation}
\eta(t) = \eta_0(t) + b_{\text{e}} e^{t\mathcal{K}_0} (3\kB  Tx_0/a_{\text{e}}-x_0^3).
\end{equation}

The memory kernel still fulfills the second fluctuation-dissipation theorem
\begin{align}
K(t,x_0(t)) &= \langle v_0^2\rangle ^{-1} \langle (\eta(0) + \mathcal{P}_\mathcal{E} (\mathcal{K}_0+\mathcal{K}_1) v_0) \eta_0(t) \rangle \nonumber \\ &= (\kB T)^{-1} \langle \eta(0) \eta(t)\rangle,
\end{align}
and thus we finally obtain
\begin{eqnarray}
K(t) = K_0(t) + \frac{b_{\text{e}}^2}{\kB  T} \langle \tilde{x}_0 (t) \tilde{x}_0 (0) \rangle_0 + \mathcal{O}(b^3)
\end{eqnarray}
with $\tilde{x}_0(t)=3\kB T x_0(t)/a_{\text{e}} - x_0(t)^3$ and where $\langle...\rangle_0$ denotes an equilibrium average in the linear system with $b_{\text{e}}=0.$  This equation therefore predicts the renormalized memory kernel of the non-linear system obtained just from quantities in the linear system and it corresponds to Eq.~(\ref{eq:Kn_smallb}) in the main text.

\noindent\textbf{Non-linear projection, non-linear potential:}
For the non-linear projection we now choose $\mathcal{I}={1,2}$ with $h_1(x_0) = x_0$, $h_2(x_0) = x_0^3$ for which we have derived the potential of mean force above. The non-linearities in the external potential $V_{\text{e}}(x_0)$ directly enter the Fokker-Planck operator,
\begin{align}\label{key}
\mathcal{K} &= v_0 \frac{\partial }{\partial x_0}  - (a_{\text{e}} x_0 + b_{\text{e}} x_0^3) \frac{\partial }{\partial v_0} + \sum_{i>0} a_i \Delta x_i \frac{\partial }{\partial v_0}\nonumber\\ &+ \sum_{i>0}  v_i \frac{\partial }{\partial x_i} - \sum_{i>0}  a_i \Delta x_i \frac{\partial }{\partial v_i} - \frac{\gamma_i}{m_i} v_i \frac{\partial }{\partial v_i} \nonumber \\&+ T \sum_{i>0} \frac{\gamma_i}{m_i^2} \frac{\partial^2 }{\partial v_i^2} .
\end{align}

To make analytical progress it is important to notice that in the definition of the noise, Eq.~(\ref{eq:defNoise}), the non-linear projection operator and the non-linear Fokker-Planck operator only come in pairs, $\mathcal{Q}_\mathcal{E} \mathcal{K}$, with $\mathcal{Q}_\mathcal{E}  = 1- \mathcal{P}_\mathcal{E} $. The non-linear term in $\mathcal{K} \propto b_{\text{e}} x_0^3 $ will therefore immediately be projected out, because, per definition, this term lies within the relevant subspace (see $h_2(x)$ above). We can therefore use the same linear ansatz as chosen in Eq.~(\ref{eq:ansatz}). In this way we explicitly find
\begin{align}
\mathcal{K} \eta(t) = &\rho_0(t) v_0 - \sigma_0(t) \left( a_{\text{e}} x_0 + b_{\text{e}}x_0^3 - \sum_{i>0}  a_i \Delta x_i \right) \nonumber\\&+ \sum_{i>0} \rho_i(t)  v_i + \sum_{i>0}\sigma_i(t) \left( a_i \Delta x_i - \gamma_i v_i \right) \nonumber\\
\mathcal{P}_\mathcal{E}\mathcal{K} \eta(t) =& \rho_0(t)  v_0 - \sigma_0(t) (a_{\text{e}} x_0+b_{\text{e}} x_0^3)
\end{align}  
where the second line follows from the same calculation as performed for the mean force. Combining the above two equations shows explicitly that the combination $\mathcal{Q}_\mathcal{E} \mathcal{K}$ is independent of the conservative force $-V_{\text{e}}^\prime(x_0)$, and thus noise $\eta(t)$ and memory kernel $K(t)$ are still given by Eqs.~(\ref{eq:FDTProject}) and (\ref{eq:MemoryProject}).

It should again be noted that the same arguments directly apply to any $V_{\text{e}}(x_0)$ up to order $ n $ in $x_0$, given that we choose $\mathcal{I} = {1,...,n-1}$ and $h_k(x_0) = x_0^k,$ even if $ n \rightarrow \infty$.

\subsection{Instantaneous memory and force correlations}
\label{app:instantaneous}

We use the theorem \cite{LandauLifshitz1969}
\begin{align}\label{eq:FNullQuadrat}
	\langle F_0(t)^2 \rangle = - \kB T \Bigl\langle \frac{\diff F_0}{\diff x_0} \Bigr\rangle.
\end{align}
This can be easily derived by writing the left-hand side as an ensemble average over the Boltzmann distribution and partial integration using the identity $ \frac{\text{d}}{\text{d} x_0} \exp(-\beta V(x_0)) = \beta  F_0 \exp(-\beta V(x_0)) $, where $V(x_0) = V_e(x_0) + \sum_i V_i(x_0-x_i) $.

Splitting $F_0(t) = -V_{\text{e}}^{\prime}(x_0(t)) + \tilde{F}(t)$ in Eq.~(\ref{eq:FNullQuadrat}) gives
\begin{align}
	\langle F_0(t)^2 \rangle & = \langle [-V_{\text{e}}^{\prime}(x_0(t)) + \tilde{F}(t)] F_0(t) \rangle \nonumber \\
	&= \langle -V_{\text{e}}^{\prime}(x_0(t)) F_0(t) \rangle + \langle \tilde{F}(t) F_0(t) \rangle \nonumber  \\
	&= - \kB T \Bigl\langle \frac{- \diff V_{\text{e}}^{\prime}(x_0(t))}{\diff x_0} + \frac{\diff \tilde{F}}{\diff x_0} \Bigr\rangle
\end{align}

Thus, $\langle \tilde{F}(t) F_0(t) \rangle = - \kB T  \bigl\langle \frac{\diff \tilde{F}}{\diff x_0} \bigr\rangle$. Inserting this and $m_0 \langle v_0^2 \rangle =  \kB T$ into the initial condition of Eq.~(\ref{eq:Kernel}) results in 
\begin{align}\label{eq:instantaneous_papa}
	K(0) = - \Bigl\langle \frac{\diff \tilde{F}}{\diff x_0} \Bigr\rangle
\end{align}

Two applications: 

We have shown in Appendix \ref{app:decomposition} that $K_{\text{e}}(t) = K_{\text{n}}(t) + \tilde{a}_{\text{e}}$. Applying this decomposition and inserting $\tilde{F} = -V_{\text{e}}^\prime$ into Eq.~(\ref{eq:instantaneous_papa}) yields $K_{\text{n}}(0) = \langle V_{\text{e}}^{\prime\prime}(x_0) \rangle - \tilde{a}_{\text{e}}$ (cf. Eq.~(\ref{eq:Kn_static2})). Since $\langle V_{\text{e}}^{\prime\prime}(x_0) \rangle = a_{\text{e}} + 3 b_{\text{e}} \langle x_0^2 \rangle$ and $\langle x_0^2 \rangle =  \kB T/\tilde{a}_{\text{e}}$, we obtain $K_{\text{n}}(0) = a_{\text{e}} - \tilde{a}_{\text{e}} + \frac{3 b_{\text{e}} \kB  T}{\tilde{a}_{\text{e}}}$ (cf. Eq.~(\ref{eq:Kn_static})).

For $\tilde{F}(x_0) = -a_i (x_0-x_i) - b_i (x_0-x_i)^3$, Eq.~(\ref{eq:instantaneous_papa}) gives $K(0) = a_i + 3b_i \langle (x_0-x_i)^2 \rangle = a_i + 3b_i  \kB T/\tilde{a}_i$ with $\tilde{a}_i =  \kB T/\langle (x_0-x_i)^2 \rangle$, cf. Eq.~(\ref{eq:KvonNull}).

\subsection{Invariants}
\label{app:invariants}

We found that the kernels do not change if $b$ and $\kB T$ are the only changing parameters and if $b \kB T$ remain constant. (Here and in the following, $b$ means $b_{\text{e}}$ and all $b_i$.) We can derive this analytically, starting from the equipartion theorem of the anharmonic potential 
\begin{align}\label{eq:invariants1}
\kB T = a \langle x^2 \rangle + b \langle x^4 \rangle
\end{align}
Here and in the following, $a$ means $a_{\text{e}}$ and all $a_i$, and $x$ means $x_0$ and all $\Delta x_i$.

Let $b^{*} = \lambda^2 b$, $\kB T^{*} = \lambda^{-2} \kB T$ ($\lambda > 0$). Then the equipartition theorem for the scaled anharmonic potential reads
\begin{align}\label{eq:invariants2}
\kB T^{*} &= a \langle {x^{*}}^2 \rangle + b^{*} \langle {x^{*}}^4 \rangle   \nonumber \\
\lambda^{-2}  \kB T &= a \langle {x^{*}}^2 \rangle + \lambda^2 b \langle {x^{*}}^4 \rangle
\end{align}
Thus, the following equation must be fulfilled:
\begin{align}\label{eq:invariants3}
a \langle x^2 \rangle + b \langle x^4 \rangle = a \lambda^2 \langle {x^{*}}^2 \rangle + b \lambda^4 \langle {x^{*}}^4 \rangle
\end{align}
That is obviously true for $x^{*} = \lambda^{-1} x$. 

By the way: From this we can conclude that $\theta = \langle x^4 \rangle/\langle x^2 \rangle ^2$ remains constant, and since $a$ and $b \kB T$ also stay constant, $\tilde{a}_{\text{e}}$ and all $\tilde{a}_i$ will not change.

Furthermore,
$m \langle v^2 \rangle =  \kB T$ and $m \langle {v^{*}}^2 \rangle =  \kB T^{*} = \lambda^{-2}  \kB T$, i.e., $m \langle v^2 \rangle = m \lambda^2 \langle {v^{*}}^2 \rangle$. This is obviously true for $v^{*} = \lambda^{-1} v$. ($m$ means here the mass of the particle $m$ and of all oscillators $m_i$, $v$ means $v_0$ and all $v_i$.)

So we found that all positional coordinates and all velocities will be scaled with the same factor, $\lambda^{-1}$, if $T$ is scaled with $\lambda^{-2}$ and all $b$ with $\lambda^2$. Thus, all frequencies will stay constant. 

Since all velocities are scaled with $\lambda^{-1}$, this will be true for $F_0$, too. For the non-linear force, this can also be shown, using the transformations derived before:
\begin{align}\label{eq:invariants3}
F^{*} &= -a x^{*} - b^{*} {x^{*}}^3   \nonumber \\
&= -a \lambda^{-1} x - \lambda^2 b \lambda^{-3} x^3   \nonumber \\
&= \lambda^{-1} (-a x - b x^3) = \lambda^{-1} F
\end{align}

Thus all correlation functions in Eq.~(\ref{eq:Kernel}) will be scaled with $\lambda^{-2}$. It is seen easily that this factor cancels throughout that equation. Therefore, all memory kernels calculated with this algorithm will not change under the assumption made above. The same is true for all $\tilde{a}$.

Along the way, we have shown above that $\theta$ remains constant for $a=$ const. and $b \kB T=$ const. However, we can show more. Let $ \kB T=$ const., $a^{*} = \nu^2 a$, $b^{*} = \nu^4 b$ ($\nu > 0$). Again we use the equipartition theorem:
\begin{align}\label{eq:invariants4}
a \langle x^2 \rangle + b \langle x^4 \rangle &=  \kB T   \nonumber \\
&= a^{*} \langle {x^{*}}^2 \rangle + b^{*} \langle {x^{*}}^4   \nonumber \rangle   \\
&= a \nu^2 \langle {x^{*}}^2 \rangle + b \nu^4 \langle {x^{*}}^4 \rangle 
\end{align}
This is obviously true for ${x^{*}} = \nu^{-1} x$. Therefore, $\theta = \langle x^4 \rangle/\langle x^2 \rangle ^2$ doesn't change if $a$ and $b$ are the only changing parameters and $\frac{a}{\sqrt{b}}$ remains constant.

We can summarize these two invariants concerning $\theta$: If $\frac{a}{\sqrt{b \kB T}} =$ const., $\theta$ will not change.

\section{Dynamics in anharmonic systems}
\label{app:anharmonic_dynamics}

\begin{figure}
	\includegraphics{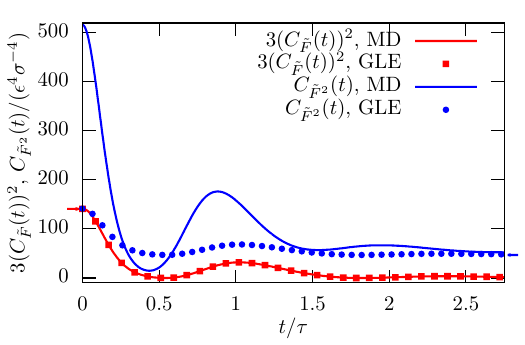}
	\caption{$C_{\tilde{F}^2}(t)$ ($= \langle \tilde{F}^2(t)\tilde{F}^2(0) \rangle$) and $3(C_{\tilde{F}}(t))^2$ ($C_{\tilde{F}}(t) = \langle \tilde{F}(t)\tilde{F}(0) \rangle$) for a free particle with a single oscillator ($a_1=0, b_1=11.42\epsilon \sigma^{-4}, m_1=1m, \gamma_1 = 6 m \tau^{-1}$). The arrow on the right-hand side shows the theoretical value of $\langle \tilde{F}^2 \rangle ^2$, the one on the left-hand side indicates $3 \langle \tilde{F}^2 \rangle ^2$ ($= \langle \tilde{F}^4 \rangle$ for a Gaussian $P(\tilde{F})$).  }
	\label{fig:c_Ftilde_cF2}
\end{figure}

In Fig.~\ref{fig:distribution_coupling} it can be seen that $P(\tilde{F})$ shows a singularity at $\tilde{F} = 0$ in the MD simulations, while it follows a Gaussian function in GLE simulations. Though these distributions are very different, their second moments ($\langle \tilde{F}^2 \rangle$) are the same since simple two-point correlations are always reproduced correctly. Nevertheless, looking at the shapes of the curves it is clear that their fourth moments cannot be identical. This implies that the autocorrelations $C_{\tilde{F}^2}(t) = \langle \tilde{F}^2(t)\tilde{F}^2(0) \rangle$ from MD and GLE simulations will not coincide. Fig.~\ref{fig:c_Ftilde_cF2} confirms these expectations, and the deviation in $C_{\tilde{F}^2}(t)$ is huge. Thus we can expect that also corresponding functions like the mean quartic displacement ($\langle (x_0(t) - x_0(0))^4 \rangle$) will not be reproduced correctly. This is true, but the differences are not very large (data not shown). In contrast, the non-Gaussian parameter $\alpha(t)$ (cf. section \ref{sec:observables}) proved to be very sensitive, which is why we focused on this function in the main text. These deviations in higher-order correlations are obviously also visible for an anharmonic external potential. There, however, they are dominated by the intrinsic differences in the distribution of positions as shown in Fig.~\ref{fig:distribution_external}b.

\FloatBarrier

\bibliography{library_local.bib}

@article{10.1063/5.0143707,
    author = {Obliger, Amaël},
    title = "{Simple and efficient algorithms based on Volterra equations to compute memory kernels and projected cross-correlation functions from molecular dynamics}",
    journal = {J. Chem. Phys.},
    volume = {158},
    number = {14},
    pages = {144101},
    year = {2023},
    month = {04},
    issn = {0021-9606},
    doi = {10.1063/5.0143707},
    url = {https://doi.org/10.1063/5.0143707},
    eprint = {https://pubs.aip.org/aip/jcp/article-pdf/doi/10.1063/5.0143707/16824800/144101\_1\_5.0143707.pdf},
}

@article{stella2014generalized,
	title={Generalized Langevin equation: An efficient approach to nonequilibrium molecular dynamics of open systems},
	author={Stella, Lorenzo and Lorenz, Chris D and Kantorovich, Lev},
	journal={Phys. Rev. B},
	volume={89},
	number={13},
	pages={134303},
	year={2014},
	publisher={APS}
}

@article{ness2017nonequilibrium,
	title={Nonequilibrium generalised Langevin equation for the calculation of heat transport properties in model 1D atomic chains coupled to two 3D thermal baths},
	author={Ness, H and Stella, L and Lorenz, CD and Kantorovich, L},
	journal={J. Chem. Phys.},
	volume={146},
	number={16},
	year={2017},
	publisher={AIP Publishing}
}

@article{vroylandt2022likelihood,
	title="{Likelihood-based non-Markovian models from molecular dynamics}",
	author={Vroylandt, Hadrien and Gouden{\`e}ge, Ludovic and Monmarch{\'e}, Pierre and Pietrucci, Fabio and Rotenberg, Benjamin},
	journal={Proc. Natl. Acad. Sci. USA},
	volume={119},
	number={13},
	pages={e2117586119},
	year={2022},
	publisher={National Acad Sciences}
}

@article{widder2022noise,
	author = {Widder,Christoph  and Koch,Fabian  and Schilling,Tanja },
	title = "{Generalized Langevin dynamics simulation with non-stationary memory kernels: How to make noise}",
	journal = {J. Chem. Phys.},
	volume = {157},
	number = {19},
	pages = {194107},
	year = {2022},
	doi = {10.1063/5.0127557},
	
	URL = { 
	https://doi.org/10.1063/5.0127557
	
	},
	eprint = { 
	https://doi.org/10.1063/5.0127557
	
	}
}

@article{maes2013fluctuation,
	title={Fluctuation-response relations for nonequilibrium diffusions with memory},
	author={Maes, Christian and Safaverdi, Soghra and Visco, Paolo and Van Wijland, Fr{\'e}d{\'e}ric},
	journal={Phys. Rev. E},
	volume={87},
	number={2},
	pages={022125},
	year={2013},
	publisher={APS}
}

@article{post2022molecular,
	title={Molecular origin of driving-dependent friction in fluids},
	author={Post, Matthias and Wolf, Steffen and Stock, Gerhard},
	journal={J. Chem. Theory Comput.},
	volume={18},
	number={5},
	pages={2816--2825},
	year={2022},
	publisher={ACS Publications}
}

@article{meyer2017non,
	title="{On the non-stationary generalized Langevin equation}",
	author={Meyer, Hugues and Voigtmann, Thomas and Schilling, Tanja},
	journal={J. Chem. Phys.},
	volume={147},
	number={21},
	pages={214110},
	year={2017},
	publisher={AIP Publishing LLC}
}

@article{zhu2022LSTM,
	author = {Zhu, Yuanran and Tang, Yu-Hang and Kim, Changho},
	year = {2022},
	archivePrefix = {arXiv},
	eprint = {2202.12278},
	journal = {arxiv:2202.12278},
	title = {Learning Stochastic Dynamics with Statistics-Informed Neural Network}
}

@article{Cairano2021_nonlinear,
	author = {Di Cairano, Loris},
	year = {2021},
	month = {12},
	pages = {},
	title = "{On the derivation of a nonlinear generalized Langevin equation}",
	volume = {6},
	journal = {J. Phys. Commun.},
	doi = {10.1088/2399-6528/ac438d}
}

@article{Xing2009_nonlinear,
	title="{Mori-Zwanzig projection formalism: from linear to nonlinear}",
	author={Xing, Jianhua},
	journal={arXiv preprint arXiv:0904.2691},
	year={2009}
}

@article{zwanzig1973nonlinear,
	title="{Nonlinear generalized Langevin equations}",
	author={Zwanzig, Robert},
	journal={J. Stat. Phys.},
	volume={9},
	number={3},
	pages={215--220},
	year={1973},
	publisher={Springer}
}

@article{Kinjo2007,
	title = {Equation of motion for coarse-grained simulation based on microscopic description},
	author = {Kinjo, Tomoyuki and Hyodo, Shi-aki},
	journal = {Phys. Rev. E},
	volume = {75},
	issue = {5},
	pages = {051109},
	numpages = {9},
	year = {2007},
	month = {May},
	publisher = {American Physical Society},
	doi = {10.1103/PhysRevE.75.051109},
	url = {https://link.aps.org/doi/10.1103/PhysRevE.75.051109}
}

@incollection{hanggi2007generalized,
	title="{Generalized Langevin equations: A useful tool for the perplexed modeller of nonequilibrium fluctuations?}",
	author={H{\"a}nggi, Peter},
	booktitle="{Stochastic Dynamics}",
	pages={15--22},
	year={2007},
	publisher={Springer}
}

@article{rudzinski2019recent,
	title={Recent progress towards chemically-specific coarse-grained simulation models with consistent dynamical properties},
	author={Rudzinski, Joseph F},
	journal={Computation},
	volume={7},
	number={3},
	pages={42},
	year={2019},
	publisher={MDPI}
}

@article{mueller2002,
	author = {Müller-Plathe, Florian},
	title = {Coarse-Graining in Polymer Simulation: From the Atomistic to the Mesoscopic Scale and Back},
	journal = {ChemPhysChem},
	volume = {3},
	number = {9},
	pages = {754-769},
	year = {2002}
}

@Article{brini2019_cg,
	author ="Brini, Emiliano and Algaer, Elena A. and Ganguly, Pritam and Li, Chunli and Rodríguez-Ropero, Francisco and van der Vegt, Nico F. A.",
	title  ="Systematic coarse-graining methods for soft matter simulations – a review",
	journal  ="Soft Matter",
	year  ="2013",
	volume  ="9",
	issue  ="7",
	pages  ="2108-2119",
	publisher  ="The Royal Society of Chemistry",
	doi  ="10.1039/C2SM27201F",
	url  ="http://dx.doi.org/10.1039/C2SM27201F"}

@article{izvekov2005multiscale,
	title={A multiscale coarse-graining method for biomolecular systems},
	author={Izvekov, Sergei and Voth, Gregory A},
	journal={J. Phys. Chem. B},
	volume={109},
	number={7},
	pages={2469--2473},
	year={2005},
	publisher={ACS Publications}
}

@article{gottwald2015_gle,
	author = {Gottwald, Fabian and Karsten, Sven and Ivanov, Sergei D. and Kühn, Oliver},
	title = "{Parametrizing linear generalized Langevin dynamics from explicit molecular dynamics simulations}",
	journal = {J. Chem. Phys.},
	volume = {142},
	number = {24},
	year = {2015},
	month = {06},
	issn = {0021-9606},
	doi = {10.1063/1.4922941},
	url = {https://doi.org/10.1063/1.4922941},
	note = {244110},
	eprint = {https://pubs.aip.org/aip/jcp/article-pdf/doi/10.1063/1.4922941/15495279/244110\_1\_online.pdf},
}

@Article{wu2018_adhesion,
	author ="Wu, Yu-Wen and Yu, Hsiu-Yu",
	title  ="{Adhesion of a polymer-grafted nanoparticle to cells explored using generalized Langevin dynamics}",
	journal  ="Soft Matter",
	year  ="2018",
	volume  ="14",
	issue  ="48",
	pages  ="9910-9922",
	publisher  ="The Royal Society of Chemistry",
	doi  ="10.1039/C8SM01579A",
	url  ="http://dx.doi.org/10.1039/C8SM01579A"}

@article{McKinley2009_diffusion,
	author = {McKinley, Scott A. and Yao, Lingxing and Forest, M. Gregory},
	title = "{Transient anomalous diffusion of tracer particles in soft matter}",
	journal = {Journal of Rheology},
	volume = {53},
	number = {6},
	pages = {1487-1506},
	year = {2009},
	month = {11},
	issn = {0148-6055},
	doi = {10.1122/1.3238546},
	url = {https://doi.org/10.1122/1.3238546},
	eprint = {https://pubs.aip.org/sor/jor/article-pdf/53/6/1487/16014115/1487\_1\_online.pdf},
}

@article{cordoba2012elimination,
	title={Elimination of inertia from a generalized Langevin equation: applications to microbead rheology modeling and data analysis},
	author={C{\'o}rdoba, Andr{\'e}s and Indei, Tsutomu and Schieber, Jay D},
	journal={Journal of Rheology},
	volume={56},
	number={1},
	pages={185--212},
	year={2012},
	publisher={The Society of Rheology}
}

@article{meyer2021_gle,
	title = {Evaluation of memory effects at phase transitions and during relaxation processes},
	author = {Meyer, Hugues and Glatzel, Fabian and W\"ohler, Wilkin and Schilling, Tanja},
	journal = {Phys. Rev. E},
	volume = {103},
	issue = {2},
	pages = {022102},
	numpages = {8},
	year = {2021},
	month = {Feb},
	publisher = {American Physical Society},
	doi = {10.1103/PhysRevE.103.022102},
	url = {https://link.aps.org/doi/10.1103/PhysRevE.103.022102}
}

@article{reith2003ibi,
	author = {Reith, Dirk and Pütz, Mathias and Müller-Plathe, Florian},
	title = {Deriving effective mesoscale potentials from atomistic simulations},
	journal = {J. Comput. Chem.},
	volume = {24},
	number = {13},
	pages = {1624-1636},
	doi = {https://doi.org/10.1002/jcc.10307},
	url = {https://onlinelibrary.wiley.com/doi/abs/10.1002/jcc.10307},
	eprint = {https://onlinelibrary.wiley.com/doi/pdf/10.1002/jcc.10307},
	year = {2003}
}

@article{noid2013_cg,
	author = {Noid, W. G.},
	title = "{Perspective: Coarse-grained models for biomolecular systems}",
	journal = {J. Chem. Phys.},
	volume = {139},
	number = {9},
	year = {2013},
	month = {09},
	issn = {0021-9606},
	doi = {10.1063/1.4818908},
	url = {https://doi.org/10.1063/1.4818908},
	note = {090901},
	eprint = {https://pubs.aip.org/aip/jcp/article-pdf/doi/10.1063/1.4818908/15465619/090901\_1\_online.pdf},
}

@book{LandauLifshitz1969,
	title={Statistical Physics},
	author={Landau, Lev Davidovich and Lifshitz, Evgenii Mikhailovich},
	year={1969},
	note={p.97, Eq. (33.14)},
	publisher={Pergamon Press}
}

@article{Daldrop2017_external,
	title = {External Potential Modifies Friction of Molecular Solutes in Water},
	author = {Daldrop, Jan O. and Kowalik, Bartosz G. and Netz, Roland R.},
	journal = {Phys. Rev. X},
	volume = {7},
	issue = {4},
	pages = {041065},
	numpages = {14},
	year = {2017},
	month = {Dec},
	publisher = {American Physical Society},
	doi = {10.1103/PhysRevX.7.041065},
	url = {https://link.aps.org/doi/10.1103/PhysRevX.7.041065}
}

@article{CLModel,
title = {Quantum tunnelling in a dissipative system},
journal = {Annals of Physics},
volume = {149},
number = {2},
pages = {374-456},
year = {1983},
issn = {0003-4916},
doi = {https://doi.org/10.1016/0003-4916(83)90202-6},
url = {https://www.sciencedirect.com/science/article/pii/0003491683902026},
author = {A.O Caldeira and A.J Leggett}
}

@article{Ceriotti2010,
	author = {Ceriotti, Michele and Bussi, Giovanni and Parrinello, Michele},
	title = {Colored-Noise Thermostats à la Carte},
	journal = {	J. Chem. Theory Comput.},
	volume = {6},
	number = {4},
	pages = {1170-1180},
	year = {2010},
	doi = {10.1021/ct900563s},
	
	URL = { 
	https://doi.org/10.1021/ct900563s
	
	},
	eprint = { 
	https://doi.org/10.1021/ct900563s
	
	}
}

@article{ayaz2021non,
	title="{Non-Markovian modeling of protein folding}",
	author={Ayaz, Cihan and Tepper, Lucas and Br{\"u}nig, Florian N and Kappler, Julian and Daldrop, Jan O and Netz, Roland R},
	journal={Proc. Natl. Acad. Sci. USA},
	volume={118},
	number={31},
	pages={e2023856118},
	year={2021},
	publisher={National Acad Sciences}
}

@article{Ayaz2022_GLENonLinear,
	title = "{Generalized Langevin equation with a nonlinear potential of mean force and nonlinear memory friction from a hybrid projection scheme}",
	author = {Ayaz, Cihan and Scalfi, Laura and Dalton, Benjamin A. and Netz, Roland R.},
	journal = {Phys. Rev. E},
	volume = {105},
	issue = {5},
	pages = {054138},
	numpages = {19},
	year = {2022},
	month = {May},
	publisher = {American Physical Society},
	doi = {10.1103/PhysRevE.105.054138},
	url = {https://link.aps.org/doi/10.1103/PhysRevE.105.054138}
}

@article{doerries2021correlation,
	title="{Correlation functions of non-Markovian systems out of equilibrium: Analytical expressions beyond single-exponential memory}",
	author={Doerries, Timo J and Loos, Sarah AM and Klapp, Sabine HL},
	journal={	J. Stat. Mech.: Theory Exp.},
	volume={2021},
	number={3},
	pages={033202},
	year={2021},
	publisher={IOP Publishing}
}

@article{Jung_2022,
	doi = {10.1088/1361-648X/ac56a7},
	url = {https://dx.doi.org/10.1088/1361-648X/ac56a7},
	year = {2022},
	month = {mar},
	publisher = {IOP Publishing},
	volume = {34},
	number = {20},
	pages = {204004},
	author = {Gerhard Jung},
	title = "{Non-Markovian systems out of equilibrium: exact results for two routes of coarse graining}",
	journal = {J. Phys.: Condens. Matter}
}

@book{Zwanzig2001,
	author = {Zwanzig, Robert},
	isbn = {0195140184},
	publisher = {Oxford University Press},
	title = {{Nonequilibrium Statistical Mechanics}},
	year = {2001}
}

@article{Zwanzig1961,
	author = {Zwanzig, Robert},
	doi = {10.1103/PhysRev.124.983},
	issn = {0031-899X},
	journal = {Phys. Rev.},
	month = {nov},
	number = {4},
	pages = {983--992},
	publisher = {American Physical Society},
	title = {{Memory Effects in Irreversible Thermodynamics}},
	url = {https://link.aps.org/doi/10.1103/PhysRev.124.983},
	volume = {124},
	year = {1961}
}

@article{Mori1965,
	author = {Mori, Hazime},
	doi = {10.1143/PTP.33.423},
	issn = {0033-068X},
	journal = {Prog. Theor. Phys.},
	month = {mar},
	number = {3},
	pages = {423--455},
	publisher = {Oxford University Press},
	title = {{Transport, collective motion, and Brownian motion}},
	url = {https://academic.oup.com/ptp/article-lookup/doi/10.1143/PTP.33.423},
	volume = {33},
	year = {1965}
}

@article{zhu2021_stochasticgle,
	author = {Zhu,Yuanran  and Venturi,Daniele },
	title = "{Hypoellipticity and the Mori–Zwanzig formulation of stochastic differential equations}",
	journal = {Journal of Mathematical Physics},
	volume = {62},
	number = {10},
	pages = {103505},
	year = {2021},
	doi = {10.1063/5.0035459},
	URL = { 
	https://doi.org/10.1063/5.0035459
	
	}
}

@article{gronbech2013simple,
	title="{A simple and effective Verlet-type algorithm for simulating Langevin dynamics}",
	author={Gr{\o}nbech-Jensen, Niels and Farago, Oded},
	journal={Molecular Physics},
	volume={111},
	number={8},
	pages={983--991},
	year={2013},
	publisher={Taylor \& Francis}
}

@article{Vroylandt2022_nonlinear,
	author = {Vroylandt,Hadrien  and Monmarché,Pierre },
	title = "{Position-dependent memory kernel in generalized Langevin equations: Theory and numerical estimation}",
	journal = {J. Chem. Phys.},
	volume = {156},
	number = {24},
	pages = {244105},
	year = {2022},
	doi = {10.1063/5.0094566},
	
	URL = { 
	 	
  https://doi.org/10.48550/arXiv.2201.02457
	
	},
	eprint = { 
	https://doi.org/10.1063/5.0094566
	
	}
	
}

@article{Shin2010,
	title = "Brownian motion from molecular dynamics",
	journal = "Chemical Physics",
	volume = "375",
	number = "2",
	pages = "316 - 326",
	year = "2010",
	note = "Stochastic processes in Physics and Chemistry (in honor of Peter Hänggi)",
	issn = "0301-0104",
	doi = "https://doi.org/10.1016/j.chemphys.2010.05.019",
	url = "http://www.sciencedirect.com/science/article/pii/S0301010410002363",
	author = "Hyun Kyung Shin and Changho Kim and Peter Talkner and Eok Kyun Lee"
}

@article{jung2017iterative,
	title={Iterative reconstruction of memory kernels},
	author={Jung, Gerhard and Hanke, Martin and Schmid, Friederike},
	journal={J. Chem. Theory Comput.},
	volume={13},
	number={6},
	pages={2481--2488},
	year={2017},
	publisher={ACS Publications}
}

@article{klippenstein2021introducing,
	title={Introducing memory in coarse-grained molecular simulations},
	author={Klippenstein, Viktor and Tripathy, Madhusmita and Jung, Gerhard and Schmid, Friederike and van der Vegt, Nico FA},
	journal={J. Phys. Chem. B},
	year={2021},
	URL = { 
	https://doi.org/10.1021/acs.jpcb.1c01120
	},
	publisher={ACS Publications}
}

@article{Rotenberg2014_inverse,
	author = {Carof,Antoine  and Vuilleumier,Rodolphe  and Rotenberg,Benjamin },
	title = {Two algorithms to compute projected correlation functions in molecular dynamics simulations},
	journal = {J. Chem. Phys.},
	volume = {140},
	number = {12},
	pages = {124103},
	year = {2014},
	doi = {10.1063/1.4868653},
	
	URL = { 
	https://doi.org/10.1063/1.4868653
	
	},
	eprint = { 
	https://doi.org/10.1063/1.4868653
	
	}
	
}

@article{barrat2011portable,
	title={Portable implementation of a quantum thermal bath for molecular dynamics simulations},
	author={Barrat, Jean-Louis and Rodney, David},
	journal={J. Stat. Phys.},
	volume={144},
	pages={679--689},
	year={2011},
	publisher={Springer}
}

@Article{bras2014_nonGaussian,
	author ="Brás, Ana R. and Gooßen, Sebastian and Krutyeva, Margarita and Radulescu, Aurel and Farago, Bela and Allgaier, Jürgen and Pyckhout-Hintzen, Wim and Wischnewski, Andreas and Richter, Dieter",
	title  ="{Compact structure and non-Gaussian dynamics of ring polymer melts}",
	journal  ="Soft Matter",
	year  ="2014",
	volume  ="10",
	issue  ="20",
	pages  ="3649-3655",
	publisher  ="The Royal Society of Chemistry",
	doi  ="10.1039/C3SM52717D",
	url  ="http://dx.doi.org/10.1039/C3SM52717D"}

@Article{cherstvy2019_nonGaussian,
	author ="Cherstvy, Andrey G. and Thapa, Samudrajit and Wagner, Caroline E. and Metzler, Ralf",
	title  ="{Non-Gaussian{,} non-ergodic{,} and non-Fickian diffusion of tracers in mucin hydrogels}",
	journal  ="Soft Matter",
	year  ="2019",
	volume  ="15",
	issue  ="12",
	pages  ="2526-2551",
	publisher  ="The Royal Society of Chemistry",
	doi  ="10.1039/C8SM02096E",
	url  ="http://dx.doi.org/10.1039/C8SM02096E"}

@article{Lesnicki2016_mh,
	title = {Molecular Hydrodynamics from Memory Kernels},
	author = {Lesnicki, Dominika and Vuilleumier, Rodolphe and Carof, Antoine and Rotenberg, Benjamin},
	journal = {Phys. Rev. Lett.},
	volume = {116},
	issue = {14},
	pages = {147804},
	numpages = {5},
	year = {2016},
	month = {Apr},
	publisher = {American Physical Society},
	doi = {10.1103/PhysRevLett.116.147804},
	url = {https://link.aps.org/doi/10.1103/PhysRevLett.116.147804}
}

@article{vanMegen1998_isf,
	title = {Measurement of the self-intermediate scattering function of suspensions of hard spherical particles near the glass transition},
	author = {van Megen, W. and Mortensen, T. C. and Williams, S. R. and M\"uller, J.},
	journal = {Phys. Rev. E},
	volume = {58},
	issue = {5},
	pages = {6073--6085},
	numpages = {0},
	year = {1998},
	month = {Nov},
	publisher = {American Physical Society},
	doi = {10.1103/PhysRevE.58.6073},
	url = {https://link.aps.org/doi/10.1103/PhysRevE.58.6073}
}

@article{wang2012_nonGaussian,
	author = {Wang, Bo and Kuo, James and Bae, Sung and Granick, Steve},
	year = {2012},
	month = {05},
	pages = {481-5},
	title = "{When Brownian diffusion is not Gaussian}",
	volume = {11},
	journal = {Nature Materials},
	doi = {10.1038/nmat3308}
}

@article{Flenner2005_nonGaussian,
	title = "{Relaxation in a glassy binary mixture: Mode-coupling-like power laws, dynamic heterogeneity, and a new non-Gaussian parameter}",
	author = {Flenner, Elijah and Szamel, Grzegorz},
	journal = {Phys. Rev. E},
	volume = {72},
	issue = {1},
	pages = {011205},
	numpages = {9},
	year = {2005},
	month = {Jul},
	publisher = {American Physical Society},
	doi = {10.1103/PhysRevE.72.011205},
	url = {https://link.aps.org/doi/10.1103/PhysRevE.72.011205}
}

@article{klippenstein2021_ccc,
	author = {Klippenstein, Viktor and van der Vegt, Nico F. A.},
	title = "{Cross-correlation corrected friction in (generalized) Langevin models}",
	journal = {J. Chem. Phys.},
	volume = {154},
	number = {19},
	year = {2021},
	month = {05},
	issn = {0021-9606},
	doi = {10.1063/5.0049324},
	url = {https://doi.org/10.1063/5.0049324},
	note = {191102},
	eprint = {https://pubs.aip.org/aip/jcp/article-pdf/doi/10.1063/5.0049324/15977641/191102\_1\_online.pdf},
}

@article{deichmann2018_polymer,
	author = {Deichmann, Gregor and van der Vegt, Nico F. A.},
	title = "{Bottom-up approach to represent dynamic properties in coarse-grained molecular simulations}",
	journal = {J. Chem. Phys.},
	volume = {149},
	number = {24},
	year = {2018},
	month = {12},
	issn = {0021-9606},
	doi = {10.1063/1.5064369},
	url = {https://doi.org/10.1063/1.5064369},
	note = {244114},
	eprint = {https://pubs.aip.org/aip/jcp/article-pdf/doi/10.1063/1.5064369/15552697/244114\_1\_online.pdf},
}

@article{klippenstein2022_asakura,
	author = {Klippenstein, Viktor and van der Vegt, Nico F. A.},
	title = "{Cross-correlation corrected friction in generalized Langevin models: Application to the continuous Asakura–Oosawa model}",
	journal = {J. Chem. Phys.},
	volume = {157},
	number = {4},
	year = {2022},
	month = {08},
	issn = {0021-9606},
	doi = {10.1063/5.0093056},
	url = {https://doi.org/10.1063/5.0093056},
	note = {044103},
	eprint = {https://pubs.aip.org/aip/jcp/article-pdf/doi/10.1063/5.0093056/16546992/044103\_1\_online.pdf},
}

@article{klippenstein2023_water,
	author = {Klippenstein, Viktor and van der Vegt, Nico F. A.},
	title = "{Bottom-up informed and iteratively optimized coarse-grained non-Markovian water models with accurate dynamics}",
	journal = {J. Chem. Theory Comput.},
	volume = {19},
	number = {4},
	pages = {1099-1110},
	year = {2023},
	doi = {10.1021/acs.jctc.2c00871},
	note ={PMID: 36745567},
	
	URL = { 
	https://doi.org/10.1021/acs.jctc.2c00871
	
	},
	eprint = { 
	https://doi.org/10.1021/acs.jctc.2c00871
	
	}
	
}

@Article{jung2018_GLD,
	author ="Jung, Gerhard and Hanke, Martin and Schmid, Friederike",
	title  ="{Generalized Langevin dynamics: construction and numerical integration of non-Markovian particle-based models}",
	journal  ="Soft Matter",
	year  ="2018",
	volume  ="14",
	issue  ="46",
	pages  ="9368-9382",
	publisher  ="The Royal Society of Chemistry",
	doi  ="10.1039/C8SM01817K",
	url  ="http://dx.doi.org/10.1039/C8SM01817K"}

@article{karniadakis2017_melt,
	author = {Li,Zhen  and Lee,Hee Sun  and Darve,Eric  and Karniadakis,George Em },
	title = "{Computing the non-Markovian coarse-grained interactions derived from the Mori–Zwanzig formalism in molecular systems: Application to polymer melts}",
	journal = {J. Chem. Phys.},
	volume = {146},
	number = {1},
	pages = {014104},
	year = {2017},
	doi = {10.1063/1.4973347},
	
	URL = { 
	https://doi.org/10.1063/1.4973347
	
	},
	eprint = { 
	https://doi.org/10.1063/1.4973347
	
	}
	
}

@article{PhysRevE.75.051109,
	title = {Equation of motion for coarse-grained simulation based on microscopic description},
	author = {Kinjo, Tomoyuki and Hyodo, Shi-aki},
	journal = {Phys. Rev. E},
	volume = {75},
	issue = {5},
	pages = {051109},
	numpages = {9},
	year = {2007},
	month = {May},
	publisher = {American Physical Society},
	doi = {10.1103/PhysRevE.75.051109},
	url = {https://link.aps.org/doi/10.1103/PhysRevE.75.051109}
}

@article{Vroylandt_2022,
	doi = {10.1209/0295-5075/acab7d},
	url = {https://dx.doi.org/10.1209/0295-5075/acab7d},
	year = {2022},
	month = {dec},
	publisher = {EDP Sciences, IOP Publishing and Società Italiana di Fisica},
	volume = {140},
	number = {6},
	pages = {62003},
	author = {Hadrien Vroylandt},
	title = "{On the derivation of the generalized Langevin equation and the fluctuation-dissipation theorem}",
	journal = {Europhys. Lett.}
}

@article{Glatzel_2021,
	doi = {10.1209/0295-5075/ac35ba},
	url = {https://dx.doi.org/10.1209/0295-5075/ac35ba},
	year = {2022},
	month = {feb},
	publisher = {EDP Sciences, IOP Publishing and Società Italiana di Fisica},
	volume = {136},
	number = {3},
	pages = {36001},
	author = {Fabian Glatzel and Tanja Schilling},
	title = {The interplay between memory and potentials of mean force: A discussion on the structure of equations of motion for coarse-grained observables},
	journal = {Europhys. Lett.}
}

@article{SciPostPhys.13.4.078,
	title = {Dynamics of a colloidal particle coupled to a Gaussian field: from a confinement-dependent to a non-linear memory},
	pages = {078},
	author = {Basu, Urna and Démery, Vincent and Gambassi, Andrea},
	journal = {SciPost Phys.},
	volume = {13},
	year = {2022},
	publisher = {SciPost},
	doi = {10.21468/SciPostPhys.13.4.078},
	url = {https://scipost.org/10.21468/SciPostPhys.13.4.078}
}

\end{document}